\documentclass[12pt]{iopart}
\usepackage[dvips]{graphicx}
\usepackage{iopams}
\usepackage{setstack}
\newcommand{\bra}[1]{\langle \, #1 \, |}
\newcommand{\ket}[1]{| \, #1 \, \rangle}
\newcommand{\elm}[3]{\bra{#1}\,#2\,\ket{#3}}
\newcommand{\id}{\hat{\mathbf{1}}}
\newcommand{\wek}[1]{{\vec{\bf {#1}}}}
\newcommand{\hhb}[1]{\fontfamily{phv}\fontseries{b}
                     \selectfont{#1}\normalfont}
\newcommand{\hopb}[1]{{\!\mbox{\hhb{#1}}}}
\newcommand{\hopbs}[1]{{\!\mbox{\hhb{#1}}}^{\,\dagger}}

\begin{document}

\title[Alternative representation of $N \times N$
density matrix]
{Alternative representation of $N \times N$ density matrix}
\author{S Kryszewski, M Zachcia{\l}}
\address{Institute of Theoretical Physics and
         Astrophysics, University od Gda{}\'{n}sk,
         ul. Wita Stwosza 57, 80-952 Gda{\'{n}}sk, Poland.}
\ead{fizsk@univ.gda.pl, dokmz@univ.gda.pl}
\begin{abstract}
We use polarization operators known from quantum theory
of angular momentum to expand the $N \times N$ dimensional
density operators. Thereby, we construct generalized Bloch
vectors representing density matrices. We study their
properties and derive positivity conditions for any $N$.
We also apply the procedure to study Bloch vector space
for a qubit and a qutrit.
\end{abstract}
\pacs{02.10.Yn, 03.65.-w}
\submitto{\JPA}


\section{Introduction}

Recent advances in the fundamentals of quantum mechanics
and in quantum information theory \cite{keyl} resulted
in the renewed interest in the properties and structure
of the space of $N \times N$ dimensional density operators
(matrices). Moreover, the composite (multipartite) systems
exhibit the effect of entanglement which still makes things
more interesting.

In order to facilitate further discussion, we briefly
summarize the essential properties of $N \times N$
dimensional density matrices. The literature on the
subject is so large, that we quote only two books
\cite{cohen,puri}. In the typical quantum-mechanical context
the $N-$level system is associated with the Hilbert space
$\mathcal{H}_{N}$ which is $N-$dimensional.
Quantum-mechanical observables are self-adjoint operators in
Banach space $\mathcal{B}(\mathcal{H}_{N})$ and may be
represented by space $\mathcal{M}_{N \times N}$ of $N \times N$
hermitian matrices. Density operators $\hat{\rho}$ for $N-$level
systems are a subset of $\mathcal{B}(\mathcal{H}_{N})$ which may
be denoted as $\mathcal{B}_{\rho} \bigl( \mathcal{H}_{N} \bigr)$.
A~density operator $\hat{\rho}\in{\cal{B}_{\rho}}({\cal{H}_{N}})$
possesses three fundamental properties
\numparts {\label{ro}}
\begin{eqnarray}
    ~~\mathrm{(i)} \quad
    & \hat{\rho} ~=~ \hat{\rho}^{\dagger},
    & \quad\mbox{-- hermiticity};
\label{roa} \\
    ~\mathrm{(ii)} \quad
    & \Tr \left\{ \, \hat{\rho} \, \right\} ~=~ 1,
    & \quad\mbox{-- normalization};
\label{rob} \\
    \mathrm{(iii)} \quad
    & \lambda_{j} \ge 0,
    & \quad\mbox{-- positivity}.
\label{roc} \end{eqnarray} \endnumparts
where $\lambda_{j}$ - eigenvalues of density operator
$\hat{\rho}$. Strictly speaking we should say that density
operators must be positive semidefinite. However, the phrase
"positive" is shorter, so we will use it, keeping in mind
the strict sense. The set ${\cal{B}_{\rho}}({\cal{H}_{N}})$
is convex: that is, if $\hat{\rho}_{1}, \hat{\rho}_{2} %
\in \mathcal{B}_{\rho} ( \mathcal{H}_{N} )$, then
\begin{equation}
   \hat{\rho}
    = s \: \hat{\rho}_{1} + ( 1 - s) \: \hat{\rho}_{2}
   \in {\cal{B}}_{\rho}({\cal{H}}_{N}),
   \hspace*{6mm} \mbox{for} \hspace*{6mm}
   s \in ( 0, \: 1 ).
\label{con01} \end{equation}
Moreover, density matrices satisfy the inequalities
\begin{equation}
    \Tr \bigl \{ \hat{\rho}^{\: k} \bigr\} \leq 1,
    \hspace*{20mm}
    \Tr \bigl \{ \hat{\rho}^{\: k} \bigr\}
    \geq \frac{1}{N^{k-1}},
\label{Tr02} \end{equation}
for $k \geq 2$, while for $k=1$ relation \eref{rob} holds.
For pure states $\hat{\rho}=\ket{\Psi} \bra{\Psi}$,
the first inequality becomes equality (the second one is
obviously satisfied). Pure states are also extremal -- they
can not be given as a nontrivial convex combination of
two other matrices.
For maximally mixed states one can find a representation
in which the corresponding matrix is diagonal with all nonzero
elements equal to $1/N$. In this case one has
$\Tr\bigl\{\hat{\rho}^{\: k}\bigr\}=1/N^{k-1}$.
Finally it may be worth noting that for two-level system
condition \eref{roc} and $\Tr\bigl\{\hat{\rho}^{\: 2}\bigr\}%
\leq 1$, are equivalent. But it is not the case  for systems
with higher dimensions, that is for $N \geq 3$. This is due
to additional conditions which follow from requirement
\eref{roc}.

Although the outlined fundamental properties of density
matrices are simple and well-known, not much is known about
the structure of the set ${\cal{B}}_{\rho}({\cal{H}}_{N})$.

Only the case of $N=2$ (also called a qubit) seems to be
an exception. One can find a one-to-one correspondence between
all two dimensional $\hat{\rho}$'s and the set od 3-dimensional
real vectors which are called Bloch vectors. To find such
a correspondence one usually uses standard Pauli matrices
$\wek{\boldsymbol{\sigma}} = ( \sigma_{1}, \;\sigma_{2},\; %
\sigma_{3} )$ which, together with a $2 \times 2 $ unit
matrix $\id_{2}$, form an orthogonal basis in
${\cal{B}}({\cal{H}}_{N})=\mathcal{M}_{2 \times 2}$.
Then one takes
\renewcommand{\arraystretch}{1.5}
\begin{equation}
  \hat{\rho}
  = \frac{1}{2} \,
    \bigl( \: \id_{2} + \wek{b} \cdot
       \wek{\boldsymbol{\sigma}} \: \bigr)
  = \frac{1}{2}
    \left( \begin{array}{cc}
         1 + b_{3}~~  & ~~b_{1} - i b_{2} \\
         b_{1} + i b_{2}~~ & ~~1 - b_{3} \end{array}
     \right),
\renewcommand{\arraystretch}{1}
\label{rob1} \end{equation}
where $\wek{b}=(b_{1}, \;b_{2},\:b_{3})\in\mathbb{R}^{3}$,
is a Bloch vector and $\wek{b} \cdot \wek{\sigma}
=b_{1}\hat{\sigma_{1}} + b_{2} \hat{\sigma_{2}} + %
b_{3}\hat{\sigma_{3}}$.
Density matrix \eref{rob1} is clearly hermitian and properly
normalized. Requirement of positivity is equivalent to the
condition $\Tr \{ \hat{\rho}^{~2} \} \leq 1 $ which yields
the inequality which must be satisfied by length
of the Bloch vector
\begin{equation}
  \bigl| \, \wek{b} \, \bigr|
  ~=~ \sqrt{\: b_{1}^{2} \:
       + \: b_{2}^{2} \: +\: b_{3}^{2}\:}
  ~\le~ 1.
\label{blo1} \end{equation}
Thus the set of $2 \times 2$ density matrices coincides
with the Bloch ball $\{\wek{b} \in \mathbb{R}^{3} : |\wek{b}|%
\leq 1\}$ of unit radius. Equality occurs only for pure states
which, therefore lie on the Bloch sphere $\{\wek{b} \in
\mathbb{R}^{3} :|\wek{b}| = 1\}$. The  sphere is a boundary
of the ball and the pure states are extremal. For maximally
mixed states one has $\bigl|\,\wek{b}\,\bigr|=0$.

On the other hand, even for $N=3$ the situation is not that
simple. In two recent papers by Kimura \cite{kim} and by Byrd
and Khaneja \cite{bkh} the $SU(N)$ generators are used
to expand an arbitrary $N \times N$ density operator.
Such generators correspond to the Pauli and Gell-Mann matrices
for $N=2,3$, respectively. The methods of construction of such
matrices is outlined in \cite{kim} (see also \cite{ts1,alen}).
The expansion coefficients form a generalized Bloch vector
which consists of $N^{2}-1$ real parameters (clearly, such
a procedure is a generalization of the $N=2$ case, because
Pauli matrices are $SU(2)$ generators). Then a one-to-one
correspondence between density operators and the allowed
generalized Bloch vectors is established. Moreover, it is shown
that due to the requirement $\Tr \{ \hat{\rho}^{2} \} \leq 1$,
the generalized Bloch vectors lie within a certain hyperball
with a finite radius. However, it is also known \cite{keyl}
that such a hyperball contains Bloch vectors corresponding
to non-positive matrices. This clearly indicates that the
structure of the set of all allowed Bloch vectors (and
therefore of all $N \times N$ density matrices) is not
simple, not to say quite complicated. Investigations of the
geometry of the space of density matrices are, therefore,
difficult and complex. An example of such studies can,
for instance, be found in \cite{jaks}.

The main aim of this work is to propose a new parametrization
of the set of $N \times N$ dimensional density matrices.
Our work is somewhat similar in its spirit to the papers
\cite{kim,bkh}, although we use other operators as a basis
in ${\cal{B}}({\cal{H}}_{N})$. We will try to argue that
the proposed representation might be useful in future
applications

The following section is devoted to the brief outline of the
results obtained via the standard $SU(N)$ generators by Kimura
\cite{kim} and by Byrd with Khaneja \cite{bkh}. We review these
results in order to compare them to the ones presented in this
work, and to argue that that the latter ones may be advantageous.

In the third section we recall the concept of polarization
operators and summarize their properties. We follow the
terminology and notation used in the handbook \cite{var}.
Similar operators (although in a less transparent notation)
are also discussed in the classic book by Biederharn and
Louck \cite{biel}. These authors also outline the application
of the polarization operators to the expansion of density
operators. We elaborate on the idea in the fourth section.
Moreover, we discuss some properties of the expansion
which seem not to be documented in the literature. Namely,
we study the analog of the generalized Bloch vector
in the light of the conditions imposed upon a density
operator.

Investigations of positivity of the density operator are
much more involved than checking hermiticity or normalization.
Hence, the fourth section is devoted to this issue.
We derive general expressions which allow construction
of positivity conditions for any $N$.

In the two next sections we employ the developed procedure
for a qubit ($N=2$) and for a qutrit ($N=3$). The first case
is simple, while in the second one we show that the positivity
requirements are quite restrictive. We study two-dimensional
cross sections of the space of generalized Bloch vectors
which possess a complicated and asymmetric structure.

Finally, in the last section we give some concluding remarks
and indicate some possible future applications and further
developments.

\section{Standard $SU(N)$ representation}

The authors of the papers \cite{kim,bkh} present very similar
ideas using, however, somewhat different notation. Their results
may be viewed as a generalization of the $N=2$ case to higher
dimensions. The idea is to expand the density operator in some,
suitably chosen basis of orthogonal (and traceless) matrices.
Such a  basis in $\mathcal{B}(\mathcal{H}_{N})$ is given by
the set of matrices which are the standard generators of the
$SU(N)$ group together with the unit matrix $\id_{N}$.
The $SU(N)$ generators  have the following properties
(for $i,k=1,2,...,N^{2}-1$)
\numparts {\label{sug}}
\begin{eqnarray}
    ~~\mathrm{(i)} \quad
    & \hat{\Lambda}_{k} = \hat{\Lambda}_{k}^{\dagger},
    & \quad\mbox{-- hermiticity};
\label{suga} \\
    ~\mathrm{(ii)} \quad
    & \Tr \left\{ \, \hat{\Lambda}_{k} \, \right\} = 0,
    & \quad\mbox{-- tracelessness};
\label{sugb} \\
    \mathrm{(iii)} \quad
    & \Tr \left\{ \, \hat{\Lambda}_{i} \,
      \hat{\Lambda}_{k} \right\} = 2 \: \delta_{ik},
    & \quad\mbox{-- orthogonality}.
\label{sugc} \end{eqnarray} \endnumparts
Property \eref{sugb} entails easy normalization of the
density operator, while (iii) (being the Hilbert-Schmidt scalar
product) ensures that the given matrices indeed form orthogonal
basis in $\mathcal{B} \bigl( \mathcal{H}_{N} \bigr)$.

The commutators and anticommutators are given as (summation
rule holds):
\numparts \label{suc}
\begin{eqnarray}
  \bigl[ \, \hat{\Lambda}_{j}, ~\hat{\Lambda}_{k} \, \bigr]
  &=& 2 \: i \: f_{jkm} \: \hat{\Lambda}_{m},
\label{suca} \\
  \bigl[ \, \hat{\Lambda_{j}}, ~\hat{\Lambda}_{k} \, \bigr]_{+}
  &=& \frac{4\:\delta_{jk}}{N} \: \id_{N}
  ~+~ 2\,g_{jkm} \, \hat{\Lambda}_{m},
\label{sucb} \end{eqnarray} \endnumparts
where $f_{jkm}$ is a completely antisymmetric tensor, while
$g_{jkm}$ is a completely symmetric tensor. As mentioned,
construction methods \cite{kim,ts1,alen} of matrices
$\Lambda_{j} - SU(N)$ generators and of the structure constants
$f_{jkm}, \:g_{jkm}$ are known, but fairly complicated.

Then, following Kimura \cite{kim} one uses operators
$\Lambda_{j}$ in ${\cal{B}}({\cal{H}}_{N})$) one can write for
the density operator (Byrd and Khaneja \cite{bkh} adopt slightly
different normalization of $b_{j}$ coefficients):
\begin{equation}
   \hat{\rho} ~=~ \frac{1}{N} \: \id_{N}
   ~+~ \frac{1}{2} \sum_{j=1}^{N^{2}-1} b_{j} \:
       \hat{\Lambda}_{j},
\label{rosu} \end{equation}
where $b_{j} \in \mathbb{R}$, give the generalized Bloch vector
$\wek{b} =\bigl(b_{1}, \: b_{2}, \ldots \ldots, b_{N^{2}-1} %
\bigr)\in \mathbb{R}^{N^{2}-1}$. Due to the properties
\eref{suga} and \eref{sugc} one sees that this expansion
clearly yields a hermitian and normalized matrix. Moreover,
requirement $\Tr\left\{ \, \rho^{2} \, \right\} \le 1$
implies that the vector $\wek{b}$ must satisfy the condition
\begin{equation}
   \bigl| \, \wek{b} \, \bigr|
   ~\equiv~ \sqrt{\sum_{j=1}^{N^{2}-1} b_{j}^{2}~}
   ~\le~ \sqrt{\frac{2(N-1)}{N}~}.
\label{llam} \end{equation}
So we see that vector $\wek{b}$ characterizing an arbitrary
$N$-dimensional density operator must lie within
($N^{2}-1$)-dimensional sphere specified by the
inequality (\ref{llam}).

As already mentioned above, for cases where $N \geq 3$
requirement \eref{Tr02} isn't equivalent to
$\lambda_{j} \geq 0$. Positivity imposes some additional
restrictions on vector $\wek{b}$. Due to that, generalized
Bloch vectors constitute a subset inside the hypersphere.
This set has complicated and asymmetric structure as briefly
discussed in \cite{kim}. Positivity of density operator is
then investigated along the same lines as we are employing
in this work. Hence, we postpone the discussion of this
issue to subsequent sections.

\section{Alternative representation for Bloch vector }

\subsection{Polarization operators}

In this section we recall the facts given in the handbook
\cite{var} (chapter 2.4.) on quantum theory of angular momentum.
For sake of completeness of this paper we define the concepts,
introduce notation and briefly give some additional
comments avoiding further reference to the given source.

Let ${\cal E}_{j}$ denote, for a given but fixed value of $j$,
the space spanned by eigenvectors $\ket{j \: m}$ of angular
momentum operator.  Number $j$ is half-integer or integer, so
it can take values: $ j= \frac{1}{2}, 1, \frac{3}{2},2, \dots $,
while obviously $m = j,j-1,\dots,-j+1,-j$.
Since for a given $j$ there are $2j+1$ vectors $\ket{j \: m}$,
the space ${\cal E}_{j}$ is $(2j+1)$-dimensional.
Vector $\ket{j \: m}$ is then represented by a column vector
of $2j+1$ components, $2j$ of them being zeroes while the
$m$-th one ($m=j,j-1,\ldots,-j$, in that order) is equal to 1.
From now on we will identify ${\cal E}_{j} = \mathcal{H}_{N} = %
\mathcal{H}_{2j+1}$, thus specifying Hilbert space for
a $N=2j+1$--level system.

In the operator space $\mathcal{B}\bigl(\mathcal{H}_{2j+1}\bigr)$
we introduce the following set of operators:
\begin{equation}
   \hopb{T}_{LM}(j)
   ~=~\sqrt{\frac{2L+1}{2j+1}\;}
      \sum_{m,m'} C^{jm}_{j m', \; LM}
      \: \ket{jm} \bra{jm'},
\label{def} \end{equation}
which  are called polarization operators.
$C^{jm}_{j m', \; LM}$ are Clebsch-Gordan coefficients (CGC).
Due to the properties of CGC one immediately sees that numbers
$L$ and $M$ are always integers and take the following values:
\begin{equation}
   L = 0,\:1,\:2, \:\ldots\ldots,\: 2j,
   \qquad
   M = -L, \: \ldots\ldots \:L,
\label{llmm} \end{equation}
It is straightforward to see that there are
$\sum_{L=0}^{2j}(2L+1) = (2j+1)^{2}$ polarization operators.
This suggests that these operators constitute the basis in
$\mathcal{B} \bigl( \mathcal{H}_{N} \bigr)$. After discussing
the properties of polarization operators, we will show
that this is indeed the case. We also note that CGC are
well-known, fully documented and easily computed by the
computer. We proceed with listing the fundamental
properties of the introduced polarization operators.

Employing the symmetry properties of CGC one easily shows
that the hermitian conjugate of operator $\hopb{T}_{LM}(j)$
is given as
\begin{equation}
   \hopbs{T}_{LM}(j) ~=~ (-1)^{M} \: \hopb{T}_{L-M}(j),
\label{therm} \end{equation}
so, in general, polarization operators are nonhermitian.
As we will see later this does not pose any serious
difficulties. On the other hand, operators $\hopb{T}_{L0}(j)$
are diagonal and hermitian.

For any $j \geq \frac{1}{2}$ and $L=0, ~M=0$, one has
\begin{eqnarray}
   \hopb{T}_{00}(j)
    ~=~\frac{1}{\sqrt{2j+1}} \; \id_{N=2j+1},
\label{too} \end{eqnarray}
so operator $\hopb{T}_{00}(j)$ is always proportional to
identity while  multiplication coefficient depends on $j$.

Operators $\hopb{T}_{LM}$ are traceless, in the sense that
\begin{equation}
    \Tr \bigl\{ \hopb{T}_{LM}(j) \bigr\}
   ~=~\sqrt{2j+1\;} \; \delta_{L0} \: \delta_{M0}.
\label{sladt} \end{equation}
Indeed, directly from the definition \eref{def} we have
\begin{equation}
   \Tr \left\{ \hopb{T}_{LM}(j) \right\}
   ~=~\sqrt{\frac{2L+1}{2j+1}\;}
      \sum_{m} C^{jm}_{j m, \; LM}
   ~=~\sqrt{\frac{2L+1}{2j+1}\;} \: \delta_{M0} \:
      \sum_{m} C^{jm}_{j m, \; L0},
\label{ttr1} \end{equation}
which follows from the condition $M+m=m$ which must be
satisfied by nonvanishing CGC. Performing the last sum one
obtains relation \eref{sladt}.

Computation of the trace of two polarization operators is more
tedious. Using the definition \eref{def} and employing the
known expressions for the sums of products of CGC, one gets
\begin{equation}
    \Tr \bigl\{ \hopb{T}_{L_{1}M_{1}}(j)
                \hopb{T}_{L_{2}M_{2}}(j) \bigl\}
   ~=~ (-1)^{M_{1}} \: \delta_{L_{1}L_{2}} \:
       \delta_{M_{1},-M_{2}}
\label{trtt} \end{equation}
From this relation and due the property \eref{therm} the
Hilbert-Schmidt product of polarization operators is given as
\begin{equation}
    \Tr \bigl\{ \hopb{T}_{L_{1}M_{1}}^{\dagger}(j)
                \hopb{T}_{L_{2}M_{2}}(j) \bigl\}
   ~=~ \: \delta_{L_{1}L_{2}} \:
       \delta_{M_{1}M_{2}}
\label{trhs} \end{equation}
From this we conclude (as it was also in the case of standard
$SU(N)$ representation) that operators $\hopb{T}_{LM}$
form an orthonormal basis in space
$\mathcal{B}(\mathcal{H}_{N=2j+1})$.

Similar calculations can be performed to obtain the traces
of multiple products of polarization operators.
Then for any $n$ one finds
\numparts \label{trnn}
\begin{eqnarray}
 & \hspace*{-25mm}
 \Tr \bigl\{ \hopb{T}_{L_{1}M_{1}}(j)
             \hopb{T}_{L_{2}M_{2}}(j) \ldots\ldots
             \hopb{T}_{L_{n}M_{n}}(j) \bigl\} ~=
\nonumber \\
 & \hspace*{-20mm}
   =~\Pi_{12\ldots n} ~\sum_{m}
   ~C^{jm+\mu_{1}}_{L_{1}M_{1}, \; jm}
   ~C^{jm+\mu_{2}}_{L_{2}M_{2}, \; jm+\mu_{1}} \ldots
    C^{jm+\mu_{n-1}}_{L_{n-1}M_{n-1}, \; jm+\mu_{n-2}}
   ~C^{jm}_{L_{n}M_{n}, \; jm+\mu_{n-1}}.
\label{trnna} \end{eqnarray}
where we have introduced the following notation
\begin{equation}
  \Pi_{12\ldots n}
  = \sqrt{\frac{(2L_{1}+1)\ldots\ldots(2L_{n}+1))}
                 {(2j+1)^{n}}}
  \qquad\mbox{and}
  \quad   \mu_{k} = \sum_{i=1}^{k} M_{i}.
\label{trnnb} \end{equation} \endnumparts
It is perhaps worth noting that  the properties of CGC imply
that $\sum_{i=1}^{n} M_{i} = 0$, if otherwise then the trace
is zero. It can be seen from the last of CGCs in \eref{trnna}
where $M_{n}+\mu_{n-1}=0$. Certainly, relations \eref{sladt}
and \eref{trtt} follow from the general expressions
\eref{trnn}. The traces of products of polarization operators
are expressed by a relatively simple formula. This should
be confronted with results of Byrd and Khaneja \cite{bkh}
who find quite involved expressions for the traces of the
products of $\hat{\Lambda}_{i}$ matrices up to $n=9$.
In the present case, formula \eref{trnna} is valid for any $n$.

The product of two polarization operators is computed in the
similar manner. Adapting the known sums of CGC to the present
needs, we express the product by a combination of polarization
operators
\begin{eqnarray} \fl
    \hopb{T}_{L_{1}M_{1}} \bigl( j \bigr)
    \hopb{T}_{L_{2}M_{2}} \bigl( j \bigr)
    &= \sum_{L_{3}} \bigl( -1 \bigr)^{2j + L_{3}}
        \sqrt{\bigl( 2L_{1}+ 1\bigr) \bigl( 2L_{2}+1 \bigr)}
        \left\{  \begin{array}{ccc} L_{1} & L_{2} & L_{3} \\
        j & j & j  \end{array} \right\}
\nonumber \\
    & \times ~C^{L_{3} M_{3}}_{L_{1}M_{1}, \; L_{2}M_{2}} \:
      \hopb{T}_{L_{3} M_{3}}(j),
\label{prtt} \end{eqnarray}
which involves Racah $6j$-coefficient. This relation can also
be used to find trace of the product of two polarization
operators. Since $\hopb{T}_{L_{3} M_{3}}(j)$ are traceless
for $L_{3} \geq 0$, only $L_{3}=M_{3}=0$ contribute to the sum.
Then $C^{0 \:0}_{L_{1}M_{1}, \; L_{2}M_{2}}$ implies that
$M_{1}=-M_{2}$ and the properties of the remaining coefficients
yield relation \eref{trtt}. Equartion \eref{prtt} allows
extension to multiplication of any number of $\hopb{T}_{LM}(j)$
operators in an easy but somewhat arduous way.

Another essential properties of any set of operators are given
by their  commutators and anticommutators. They follow
immediately from expression \eref{prtt} and from symmetry
properties of CGC, and are given as:
\begin{eqnarray}  && \fl
   \bigl[ \hopb{T}_{L_{1}M_{1}}(j),
   ~\hopb{T}_{L_{2}M_{2}}(j) \bigl]_{\pm}
    =\sqrt{(2L_{1}+1)(2L_{2}+1)\:}
      \sum_{L_{3}} ~(-1)^{2j+L_{3}}
\nonumber \\
  && \times
    \bigl[ \; 1 \pm (-1)^{L_{1}+L_{2}+L_{3}} \; \bigr]
    \left\{  \begin{array}{ccc} L_{1} & L_{2} & L_{3} \\
    j & j & j  \end{array} \right\}
    ~C^{L_{3}M_{3}}_{L_{1}M_{1}, \; L_{2}M_{2}} \:
    \hopb{T}_{L_{3}M_{3}}(j),
\label{cctt} \end{eqnarray}
These relations also show that polarization operators
(slightly differently normalized) are just another
realization of the representation of the $SU(N)$ group.
The above relations specify structural constants other that
those used in \cite{kim,bkh}. This explains why we have
called the representation used in these papers "a standard
one".  It is worth stressing that properties of
$\hopb{T}_{LM}(j)$ operators follow directly from the
symmetry properties of Clebsch-Gordan coefficients and
from properties of other quantities well-known from the
impressive literature on quantum theory of angular momentum.
We also note that Biederharn and Louck give some interesting
comments on the structure of the present alternative
representation of $SU(N)$ group generated by operators
$\hopb{T}_{LM}(j)$. Discussion of these issues, although
quite interesting, goes beyond the scope of this work,
but perhaps deserves some further study.

Standard generators of $SU(N)$ formed a basis in
$\mathcal{B} \bigl( \mathcal{H}_{N} \bigr)$. The same applies
to polarization operators, therefore they can be clearly used
to expand operators. Let $\hat{A} \in \mathcal{B} %
\bigl( \mathcal{H}_{N=2j+1} \bigr)$ with number $j$
specifying the dimension of the Hilbert space: $N=2j+1$.
Since $j$ is fixed for a given physical system, henceforward
we will not write it explicitly where it is not essential.

Obviously, due to completeness of states $\ket{j\,m}$ which
span the space ${\cal H}_{N=2j+1}$ we can decompose any
operator $\hat{A}$ acting on this space as follows
\begin{equation}
    \hat{A}
    = \sum_{m} \sum_{m'} A_{mm'} \ket{jm} \bra{jm'},
\label{roza} \end{equation}
with  $A_{mm'} = \elm{jm}{\hat{A}}{jm'}$ being the
corresponding matrix element of $\hat{A}$. The set of
matrix elements $A_{mm'}$ completely describes operator
$\hat{A}$ in a given basis.

Since polarization operators $\hopb{T}_{LM}$ constitute
another orthonormal  basis in $\mathcal{B} \bigl( %
\mathcal{H}_{N} \bigr)$ we can decompose any operator in
this basis, thus writing
\begin{equation}
    \hat{A}
    = \sum_{L=0}^{2j} \sum_{M=-L}^{L}
      \widetilde{A}_{LM} \: \hopb{T}_{LM},
\label{atlm} \end{equation}
Due to orthonormality relation \eref{trhs} the  components
$\widetilde{A}_{LM}$ of the above expansion are given as
\begin{equation}
    \widetilde{A}_{LM}
   = \Tr \bigl\{ \hopbs{T}_{LM} \: \hat{A} \bigr\}.
\label{elma} \end{equation}
These coefficients completely describe operator $\hat{A}$
in the basis of $\hopb{T}_{LM}$ operators.

Employing the well-known orthogonality relations for CGC
one can find the relations between $A_{mm'}$ and $A_{LM}$
\numparts \label{t11}
\begin{eqnarray}
   \widetilde{A}_{LM}
   &= \sum_{m=-j}^{j}  \sum_{m'=-j}^{j}
        \sqrt{\frac{2L+1}{2j+1}\:}
        ~C^{jm}_{j m', \; LM} \: A_{mm'},
\label{t11a} \\
   A_{mm'}
   &= \sum_{L=0}^{2j} \sum_{M=-L}^{L}
        \sqrt{\frac{2L+1}{2j+1}\:}
        ~C^{jm}_{j m', \; LM} \: \widetilde{A}_{LM}.
\label{t11b} \end{eqnarray} \endnumparts

\subsection{Decomposition of density matrix}

Clearly, a decomposition such as discussed above may also be
applied to density operator describing the $N=2j+1$-level system
(see also chapter 7 of Reference \cite{biel}).
Since operator $\hopb{T}_{00}$ plays special role (due to the
relation \eref{too}) we will write it outside the sum.
Hence we have, for the considered density operator,
the following expansion
\begin{equation}
    \hat{\rho}
    = V_{00} \hopb{T}_{00}
      + \sum_{L=1}^{2j} \sum_{M=-L}^{M=L} V_{LM} \hopb{T}_{LM}
    = V_{00} \hopb{T}_{00}
      + \wek{V} \cdot \wek{\hopb{T}}.
\label{rhot1} \end{equation}
where $\wek{V}$ is again called a generalized Bloch vector,
similarly as it was done in case of the expansion in terms of
standard $SU(N)$ generators. This vector consists
of $N^{2}-1=(2j+1)^{2}-1$ components ordered as
\begin{equation}
    \wek{V}
    = \bigl(
        V_{1-1},V_{10},V_{11},V_{2-2},V_{2-1}  ...,V_{LM}
    \bigr).
\label{vord} \end{equation}
The  component $V_{LM}$ of the generalized Bloch vector is
given similarly as in \eref{elma}, that is
\begin{equation}
   V_{LM} = \Tr \bigl\{ \hopbs{T}_{LM}\: \hat{\rho} \bigr\}.
\label{vlma} \end{equation}
Polarization operators are traceless. This fact, together with
the normalization condition \eref{rob} and with relation
\eref{sladt}, imply that
\begin{equation}
   1 = \Tr \bigl\{ \, \hat{\rho} \, \bigr\}
     = V_{00} \; \Tr \bigl\{ \, \hopb{T}_{00} \, \bigr\}
     = V_{00} \sqrt{2j+1},
\label{voo1} \end{equation}
and therefore, for any fixed $N=2j+1$ we have
\begin{equation}
    V_{00}=\frac{1}{\sqrt{2j+1}}=\frac{1}{\sqrt{N}},
\label{voo}
\end{equation}
which is then also fixed. Hence the decomposition \eref{rhot1}
may be rewritten as
\begin{equation}
    \hat{\rho}
    = \frac{1}{N} \: \id_{N=2j+1}
      + \wek{V} \cdot \wek{\hopb{T}},
\label{rhot2} \end{equation}
where the product $\wek{V} \cdot \wek{\hopb{T}}$ is specified
by \eref{rhot1}. Polarization operators are traceless, thus,
the required normalization of the density operator is always
preserved. Moreover, this explains why we have restricted
the generalized Bloch vector $\wek{V}$ to $L \geq 1$, as
indicated in definition \eref{vord}.

Density operator must be hermitian, while polarization operators
$\hopb{T}_{LM}$ are not. Therefore, components $V_{LM}$ of the
generalized Bloch vector $\wek{V}$ are in general complex.
Due to relation (\ref{therm}) the hermiticity of the density
operator implies that the complex conjugates of components
$V_{LM}$ are given as
\begin{equation}
    V_{LM}^{\ast} ~=~ \bigl( -1 \bigr)^{M} V_{L-M}
\label{vherm} \end{equation}
Components of the generalized Bloch vector have the following
properties:
\begin{itemize} \parskip=-1mm
\item there are $N-1=2j$ components $V_{L0}$ (with $L \geq 1$).
      Since $\hopbs{T}_{L0}=\hopb{T}_{L0}$ numbers $V_{L0}$ are
      real;
\item there are $N(N-1)$ components $V_{LM}$
      (with $L \geq 1$ and $M \neq 0$).
      Relation \eref{vherm} implies that:
      (i) for even $M$, if $V_{LM} = \alpha + i \, \beta$
      (with $\alpha, \beta \in \mathbb{R}$) then
      $V_{L-M} = \alpha - i \, \beta$;
      (ii) for odd $M$, if $V_{LM} = \alpha + i \, \beta$
      (with $\alpha, \beta \in \mathbb{R}$) then
      $V_{L-M} = - \alpha + i \, \beta$. So these components
      are fully represented by $N(N-1)$ real numbers.
\end{itemize}
Therefore, we conclude that relations \eref{vherm} ensure
hermiticity of the density operator and that $\wek{V}$
is given by $N^{2}-1$ real numbers, just like the corresponding
generalized Bloch vector in standard $SU(N)$ representation.
Thus, we can say that using new type of operator basis we obtain
alternative decomposition  of $N$-dimensional density operator,
which preserves two requirements: hermiticity and normalization.
The question of positivity is much more involved, so we shall
discuss it separately in the next section.

Before doing so, we would like to add some additional remarks.
The fact that  $V_{LM}$ are complex has another interesting
consequence, namely, their values as defined in \eref{vlma}
cannot be measured directly. We have to construct a set of
hermitian observables using polarization operators.
The solution is simple, for instance, we can take
the following hermitian combinations:
\begin{eqnarray}
   \mathbf{Q}_{LM}
   &= \hopb{T}_{LM} + \hopbs{T}_{LM},
\label{t22a} \\
   \mathbf{\widetilde{Q}}_{LM}
   &= i \bigl(\hopb{T}_{LM} - \hopbs{T}_{LM} \bigr),
\label{t22b} \nonumber 
\end{eqnarray}
and now $V_{LM}$ are expressed by the expectation values of
these observables
\begin{equation}
   V_{LM}
   = \frac{1}{2}
     \bigl< \mathbf{Q}_{LM} \bigr>
     + \frac{i}{2} \bigl<\mathbf{\widetilde{Q}}_{LM} \bigr>,
\label{qqtt} \end{equation}
which, in particular, yields $V_{L0}=\bigl<\hopb{T}_{L0}\bigr>$.
Specific experimental needs may require construction of yet
anotherset of hermitian observables, however, we have shown
that this is clearly possible to do so. Hence we see that the
fact that $\hopb{T}_{LM}$ are nonhermitian does not present
any difficulty.

Moreover, for the sake of future needs, we define the scalar
product of two generalized Bloch vectors $\wek{V}_{1}$ and
$\wek{V}_{2}$ by the following formula
\begin{equation}
   \wek{V}_{1} \cdot \wek{V}_{2}
   ~=~ \sum_{L=1}^{2j} \sum_{M=-L}^{L}
        V_{LM}^{(1)} \bigl( V_{LM}^{(2)} \bigr)^{\ast}
    ~=~ \sum_{L=1}^{2j} \sum_{M=-L}^{L}
        (-1)^{M} V_{LM}^{(1)} V_{L-M}^{(2)}
\label{vvsp} \end{equation}

As a next remark we give a restriction on the length
$\bigl| \wek{V} \bigr|^{2} = \wek{V} \cdot \wek{V}$
(in the sense of \eref{vvsp}). Any density operator must
satisfy inequalities \eref{Tr02}. Employing expansion
\eref{rhot2}, using the tracelessness of polarization
operators  and expression \eref{trtt} we arrive at
the equation
\begin{eqnarray}
  \Tr \bigl\{ \: \hat{\rho}^{\, 2} \: \bigr\}
  &=& \frac{1}{N}
  + \Tr \left\{ \left(\wek{V} \cdot \wek{\hopb{T}}
     \right)^{2} \right\}
\nonumber \\
&=& \frac{1}{N}
  + \sum_{L=1}^{2j} \sum_{M=-L}^{L} V_{LM} V_{LM}^{\ast}
  = \frac{1}{N} + \bigl| \wek{V} \bigr|^{2}
  \leq 1.
\label{vvsq1} \end{eqnarray}
Note that we have $\Tr\left\{\left(\wek{V}\cdot\wek{\hopb{T}}%
\right)^{2} \right\} = \bigl| \wek{V} \bigr|^{2}$. Relation
\eref{vvsq1} yields the mentioned restriction
\begin{equation}
    \bigl| \wek{V} \bigr|
    \leq \sqrt{\frac{N-1}{N}} = \sqrt{\frac{2j}{2j+1}},
\label{vvsq2} \end{equation}
This result is analogous to relation \eref{llam}
obtained in the case of the standard $SU(N)$ representation.
Hence, we can say that that all generalized Bloch vectors
$\wek{V}$ form a subset within a hypersphere of radius
$R=\sqrt{(N-1)/N}$. The structure of this subset
(for $N \geq 3$) is, as we know, pretty complicated because
requirement of positivity imposes additional restrictions
on this subset. Nevertheless, pure states (for which
$\Tr \bigl\{ \: \hat{\rho}^{2} \: \bigr\}=1$) lie on the
surface of the hypersphere of radius $R$. On the other hand,
maximally mixed states (for which
$\Tr \bigl\{ \: \hat{\rho}^{2} \: \bigr\}=1/N$) correspond
to $\bigl| \wek{V} \bigr|=0$. We can say that the shorter is
vector $\wek{V}$, the density operator represented by it
corresponds to a "more mixed" state. The length of $\wek{V}$
can be, thus, considered as a kind of measure of "mixedness"
\cite{mix}.

Finally, we note that for two density operators $\hat{\rho}_{1}$
and $\hat{\rho}_{2}$ one has $0 \leq \Tr\{ \hat{\rho}_{1} %
\hat{\rho}_{2} \} \leq 1$ (see \cite{jaks}). Using two expansions
\eref{rhot2} and due to tracelessness of $\hopb{T}_{LM}$'s
we obtain inequalities for corresponding vectors $\wek{V}_{1}$
and $\wek{V}_{2}$, namely
\begin{equation}
   - \: \frac{1}{N}
   \leq \wek{V}_{1} \cdot \wek{V}_{2}
   \leq \frac{N-1}{N}.
\label{ort2} \end{equation}
If both these vectors describe pure states then their lengths
are equal $\sqrt{(N-1)/N}$ and the above inequalities are
expressed in terms of $\cos \alpha$ -- cosine of the angle
between two considered vectors. Thus, we obtain for this case
\begin{equation}
   - \: \frac{1}{N-1} \leq \cos \alpha \leq 1.
\label{ort1} \end{equation}
This result reproduces the ones obtained in \cite{kim} and
\cite{bkh}. It confirms the notion that the set consisting
of all allowed generalized Bloch vectors is complicated
and quite likely asymmetric.

\section{Positivity of density operator}

Any density operator, apart from being normalized and hermitian,
must also be positive. Requirement of positivity may be
formulated in many ways. For example, operator $\hat{\rho}$
is positive if
\begin{equation}
   \forall ~\ket{\psi} \in {\cal{H}_{N}}
   \hspace*{6mm} \Rightarrow \hspace*{6mm}
   \elm{\psi}{\hat{\rho}}{\psi} \ge 0.
\label{poaa} \end{equation}
Another formulation is given via the eigenvalues, as it was
stated in \eref{roc}. We shall concentrate on the
latter approach since it was also used by the authors of papers
\cite{kim,bkh}. In this manner we would be able to compare our
results with those of Kimura and Byrd with Khaneja.

Thus, we need necessary tools to investigate
the eigenvalues of the $N$-dimensional density operator.
Such a tool is provided by the characteristic polynomial
of a variable $\lambda \in \mathbb{R}$:
$W(\lambda) =   \det \bigl( \lambda \id_{N} - \hat{\rho}\bigr)$
(see Ref.\cite{gant}).
Note that, in comparison with the usual notation, we have
changed the sign. It can be shown that the polynomial
$W(\lambda)$ may be written as
\begin{equation}
  W(\lambda)
  = \sum_{k=0}^{N} (-1)^{k} S^{(N)}_{k} \lambda^{N-k},
\label{pobb} \end{equation}
with $S^{(N)}_{0} \equiv 1$. It is also worth noting that
$S^{(N)}_{N} = \det \hat{\rho}$. Coefficients $S^{(N)}_{k},%
~(k= 1,2, \dots,N)$, are constructed recursively by the
Newton's formula
\begin{equation}
  k \: S^{(N)}_{k} = \sum_{m=1}^{k} (-1)^{m-1} S^{(N)}_{k-m}
    \Tr \bigl\{ \hat{\rho}^{\, m} \bigr\}.
\label{pocc} \end{equation}
Obviously $S^{(N)}_{1} \equiv \Tr \bigl\{ \hat{\rho} \bigr\}%
= 1$ due to normalization of the density operator. Computation
of subsequent $S^{(N)}_{k}$'s is straightforward. Several
initial quantities $S^{(N)}_{k}$ are as follows
\numparts \label{sskk}
\begin{eqnarray}
  S^{(N)}_{2}
  &=& \frac{1}{2}
    - \frac{1}{2} \; \Tr \bigl\{ \hat{\rho}^{\, 2} \bigr\},
\label{sskka} \\
  S^{(N)}_{3}
  &=& \frac{1}{6}
    - \frac{1}{2} \; \Tr \bigl\{ \hat{\rho}^{\, 2} \bigr\}
    + \frac{1}{3} \; \Tr \bigl\{ \hat{\rho}^{\, 3} \bigr\},
\label{sskkb} \\
  S^{(N)}_{4}
  &=& \frac{1}{24}
    - \frac{1}{4} \; \Tr \bigl\{ \hat{\rho}^{\, 2} \bigr\}
    + \frac{1}{8} \;\left( \Tr \bigl\{ \hat{\rho}^{\, 2} \bigr\}
      \right)^{2}
    + \frac{1}{3} \; \Tr \bigl\{ \hat{\rho}^{\, 3} \bigr\}
    - \frac{1}{4} \; \Tr \bigl\{ \hat{\rho}^{\, 4} \bigr\},
\label{sskkc} \end{eqnarray} \endnumparts
and will be useful in the study of some examples in subsequent
sections. The same results (although in different notation)
are also given in Refs.\cite{kim,bkh}.
Having constructed the coefficients $S_{k}$ of the characteristic
polynomial \eref{pobb} of the density operator $\hat{\rho} \in%
{\cal B}_{\rho}({\cal H}_{N})$, we can address the question
of positivity. The answer is supplied by the following theorem
\begin{equation} \fl
    \bigl\{ S^{(N)}_{k} \geq 0, ~\mbox{ for all } k=1,2,\ldots,N
        \: \bigr\}
    \hspace*{6mm}
    \Longleftrightarrow
    \hspace*{6mm}
    \bigl\{ \forall_{j} \; \lambda_{j} \geq 0,
    ~\mbox{that is}~
    \hat{\rho} - \mbox{positive} \bigr\}.
\label{podd} \end{equation}
Hence, to check whether a given operator $\hat{\rho}$ is indeed
positive, one needs to check the positivity of the corresponding
coefficients $S^{(N)}_{k}$.
On the other hand, requirement that $S^{(N)}_{k}, %
~(k=2, \ldots, N)$ are nonnegative imposes restrictions on
the components of the vector $\wek{V}$, thereby inducing
a complex structure on the set of all allowed $\wek{V}$'s.

Certainly the condition $S^{(N)}_{2} \geq 0$
(see \eref{sskka}) implies $\Tr \{ \hat{\rho}^{2} \} \leq 1$
and therefore reproduces the requirement $0 \le |\wek{V}|^{2} %
\le (N-1)/N$, as already discussed in \eref{vvsq2}.
So, for a qubit (when $j=1/2$ so that $N=2$) the latter
requirement is indeed equivalent to the requirement of positivity.
For higher dimensions $S^{(N)}_{k}$ $(k=2, \ldots, N)$, and then
the condition imposed upon $\Tr \{ \hat{\rho}^{2} \}$ is necessary
but not sufficient to ensure positivity. The higher $S^{(N)}_{k}$
(that is for $k \geq 3$) must also be checked for nonnegativity.
This simple remark, strangely enough, seems not to be noticed in
papers \cite{kim} and \cite{bkh}.

Quantities $S^{(N)}_{k}$ are easily computed
provided the traces $\Tr \bigl\{ \hat{\rho}^{\, k} \bigr\}$
are known. Density operator is normalized and equation
\eref{vvsq1} gives $ \Tr \bigl\{ \hat{\rho}^{\,2} \bigr\} = %
1/N + \bigl| \wek{V} \bigr|^{2}$. For $k \geq 3$ we employ
expansion \eref{rhot2} which yields
\begin{equation}
     \Tr \bigl\{ \hat{\rho}^{\, k} \bigr\}
   = \Tr \left\{ \left( \frac{1}{N} \; \id_{N}
       + \wek{V} \cdot \wek{\hopb{T}}\right)^{k} \right\}.
\label{poff} \end{equation}
Using Newton's binomial and tracelessness of polarization
operators we have
\begin{eqnarray}
   \Tr \bigl\{ \hat{\rho}^{\, k} \bigr\}
   &=& \sum_{m=0}^{k}
       \renewcommand{\arraystretch}{0.7}
       \left( \begin{array}{c} k \\ m \end{array} \right)
       \frac{T_{m}}{N^{k-m}}
\nonumber \\
   &=& \frac{1}{N^{k-1}}
     + \frac{k(k-1)}{2N^{k-2}} ~\bigl|\wek{V} \bigr|^{2}
     + \sum_{m=3}^{k}
       \renewcommand{\arraystretch}{0.7}
       \left( \begin{array}{c} k \\ m \end{array} \right)
       \frac{T_{m}}{N^{k-m}},
\renewcommand{\arraystretch}{1}
\label{pogg} \end{eqnarray}
where we have denoted
\begin{equation}
    T_{m} = \Tr \left\{ \left( \wek{V} \cdot \wek{\hopb{T}}
            \right)^{m} \right\},
\label{tmmm} \end{equation}
which, due to hermiticity of $\hat{\rho}$ are real.
In expression \eref{pogg} we understand that
$(\wek{V}\cdot\wek{\hopb{T}})^{0} = \id_{N}$ which entails
that $T_{0}=N$. Moreover, tracelessness of polarization
operators implies that $T_{1}=0$ and relation \eref{vvsq1}
gives $T_{2}=|\wek{V}|^{2}$. So we can say that the problem
is now reduced to computation of the quantities $T_{m}$ for
$m \geq 3$. Directly from the definition, it follows that
one can write
\begin{equation} \fl
  T_{k}
  = \sum_{L_{1}=1}^{2j} \sum_{M_{1}=-L_{1}}^{L_{1}}
    \dots \sum_{L_{k}=1}^{2j} \sum_{M_{k}=-L_{k}}^{L_{k}}
    V_{L_{1}M_{1}} \dots V_{L_{k}M_{k}}
    \Tr \left\{ \hopb{T}_{L_{1}M_{1}}
         \dots \hopb{T}_{L_{k}M_{k}} \right\},
\label{pohh} \end{equation}
and since the multiple trace is known (see formulas \eref{trnn})
we can find any $T_{k}$ and therefore the traces
$\Tr \{ \, \hat{\rho}^{k} \}$. Resulting expressions are
complicated but the multiple trace is not zero only when
$\sum_{i=1}^{k} M_{i}=0$ which greatly reduces the
number of terms.

Before constructing explicit expressions for coefficients
$S^{(N)}_{k}$ we write down traces of $\hat{\rho}^{\,k}$
for $k = 3, 4$. They will be useful later and are as follows
\numparts \label{tr34}
\begin{eqnarray}
    \Tr \bigl\{ \rho^{3} \bigr\}
    &=& \frac{1}{N^{2}}
    + \frac{3}{N} ~ \bigl| \wek{V} \bigr|^{2} + T_{3},
\label{tr34a} \\
    \Tr \bigl\{ \rho^{4} \bigr\}
    &=& \frac{1}{N^{3}}
    + \frac{6}{N^{2}} ~ \bigl| \wek{V} \bigr|^{2}
    + \frac{4}{N} \: T_{3} + T_{4}.
\label{tr34b} \end{eqnarray} \endnumparts
Coefficients $S^{(N)}_{k}$ follow by combining the recurrence
relation \eref{pocc} with the first of equations \eref{pogg}
\begin{equation}
  n \: S^{(N)}_{n}
  = \sum_{k=1}^{n} (-1)^{k-1} \: S^{(N)}_{n-k}
    ~\sum_{m=0}^{k}
     \renewcommand{\arraystretch}{0.7}
     \left( \begin{array}{c} k \\ m \end{array} \right)
     \frac{T_{m}}{N^{k-m}}
\label{sk1} \end{equation}
Writing out the $k=1$ term, we obtain
\begin{equation} \fl
  n \: S^{(N)}_{n}
  = S^{(N)}_{n-1}~\sum_{m=0}^{1}
    \renewcommand{\arraystretch}{0.7}
    \left( \begin{array}{c} 1 \\ m \end{array} \right)
    \frac{T_{m}}{N^{1-m}}
  ~+~ \sum_{k=2}^{n} (-1)^{k-1} \: S^{(N)}_{n-k}
   ~\sum_{m=0}^{k}
    \renewcommand{\arraystretch}{0.7}
    \left( \begin{array}{c} k \\ m \end{array} \right)
    \frac{T_{m}}{N^{k-m}}.
\label{sk2} \end{equation}
Certainly, the second term contributes only for $n \geq 2$.
Since $S^{(N)}_{0} = S^{(N)}_{1} = 1$ we can safely assume
that  $n \geq 2$. Then, we note that $T_{0}=N$ and $T_{1}=0$,
hence
\begin{equation}
  n \: S^{(N)}_{n}
  = S^{(N)}_{n-1}
   + \sum_{k=2}^{n} (-1)^{k-1} \: S^{(N)}_{n-k}
     \left[ \frac{1}{N^{k-1}}
      + \sum_{m=2}^{k}
        \renewcommand{\arraystretch}{0.7}
        \left( \begin{array}{c} k \\ m \end{array} \right)
        \frac{T_{m}}{N^{k-m}}
     \right],
\label{skff} \end{equation}
which is the sought recurrence relation for coefficients
$S^{(N)}_{n}$. The traces $T_{m}$ can be computed as discussed
above. The first nontrivial coefficients are
(with $T_{2}=|\wek{V}|^{2})$)
\numparts \label{skn}
\begin{eqnarray}
  S^{(N)}_{2}
  &=& \frac{N-1}{2N}
    - \frac{1}{2} \; T_{2},
\label{skna} \\
  S^{(N)}_{3}
  &=& \frac{(N-1)(N-2)}{6N^{2}}
    - \frac{N-2}{2N} \; T_{2}
    + \frac{1}{3} \; T_{3},
\label{sknb} \\
  S^{(N)}_{4}
  &=& \frac{(N-1)(N-2)(N-3)}{24 N^{3}}
    - \frac{(N-2)(N-3)}{4 N^{2}} \; T_{2}
\nonumber \\
  &&
    + \frac{N-3}{3N} \; T_{3}
    - \frac{1}{4} \; T_{4}
    + \frac{1}{8} \; T_{2}^{2}.
\label{sknc} \end{eqnarray} \endnumparts
As the theorem \eref{podd} states, positivity of $\hat{\rho}$ is
equivalent to the conditions that $S^{(N)}_{k} \geq 0$ for all
$k=1,2, \ldots, N$. Relation \eref{skff} allows one to compute
these quantities for any finite $N$. These computations might
be lengthy or tedious, but otherwise straightforward.
This follows from expression \eref{pohh} which together
with relations \eref{trnn} allow us to compute traces $T_{k}$.
Thus, we conclude, that the proposed approach to the
parametrization of $N$-dimensional density operator
can be applied in a closed form for any $N$.
On the other hand, Kimura \cite{kim} gives specific
expressions only for $N \leq 4$ while Byrd and Khaneja
\cite{bkh} give up to $N \leq 9$. Our presentation is free
from such restrictions. We give quite specific expressions
valid for any $N$.

\section{Example: Qubit}

The formalism introduced above can now be applied in some specific
cases. The simplest one is a qubit and it corresponds to
$j=1/2$, hence to $N=2j+1=2$, which within the "standard
$SU(N)$ framework was described by Pauli matrices and Bloch
vector. In this case prescription \eref{def} yields $L=0,1$
and $M=-L,\dots,L$ for each $L$. The matrices representing
polarization operators are of the form
\begin{eqnarray}
   \hopb{T}_{00}\bigl( \frac{1}{2} \bigr)
   & = \frac{1}{\sqrt{2}}
       \renewcommand{\arraystretch}{1.0}
       \left( \begin{array}{cc} 1 & 0 \\ 0 & 1
              \end{array} \right),
   & \qquad
     \hopb{T}_{1+1}\bigl( \frac{1}{2} \bigr)
     \renewcommand{\arraystretch}{1.0}
     =- \left( \begin{array}{cc} 0 & 1 \\ 0 & 0
                \end{array} \right),
\nonumber \\[2mm]
   \hopb{T}_{10}\bigl( \frac{1}{2} \bigr)
   & = \frac{1}{\sqrt{2}}
     \renewcommand{\arraystretch}{1.0}
     \left( \begin{array}{cc} 1 & 0 \\ 0 & -1
            \end{array} \right),
   & \qquad
     \hopb{T}_{1-1}\bigl( \frac{1}{2} \bigr)
     \renewcommand{\arraystretch}{1.0}
   =~\left( \begin{array}{cc} 0 & 0 \\ 1 & 0
             \end{array} \right).
\label{tj1m} \end{eqnarray}
Then, according to \eref{rhot2} we can represent the
2-dimensional density matrix by 3-di\-men\-sio\-nal
vector $\wek{V}$,
\begin{equation}
\renewcommand{\arraystretch}{1.5}
\arraycolsep=4mm
   \hat{\rho} = \frac{1}{N} \; \id_{2}
    + \sum_{M=-1}^{1} V_{1M} \hopb{T}_{1M}
   = \left( \begin{array}{cc}
       \displaystyle{\frac{1}{2} + \frac{V_{11}}{\sqrt{2}}}
    & - V_{11}
    \\
    -V_{11}^{\ast}
     &
     \displaystyle{\frac{1}{2} - \frac{V_{11}}{\sqrt{2}}}
     \end{array} \right).
\renewcommand{\arraystretch}{1.0}
\label{qub1} \end{equation}
Denoting $V_{10}=x \in \mathbb{R}$, $V_{11}=\alpha + i \beta %
\in \mathbb{C}$ and using relation \eref{vherm} we get in terms
of real parameters
\renewcommand{\arraystretch}{1.5}
\begin{equation}
   \hat{\rho}
   = \left( \begin{array}{cc}
       \displaystyle{\frac{1}{2} - \frac{x}{\sqrt{2}}}
    ~~&~~
    - \alpha - i \beta
    \\
    -\alpha + i \beta
     ~~&~~
     \displaystyle{\frac{1}{2} + \frac{x}{\sqrt{2}}}
     \end{array} \right).
\renewcommand{\arraystretch}{1.0}
\label{qub2} \end{equation}
this matrix is clearly hermitian and normalized. The requirement
of positivity $S^{(2)}_{2} \geq 0$ is equivalent to the condition
$\Tr\{ \hat{\rho}^{2} \} \leq 1$, which then gives
\begin{equation}
    \bigl| \wek{V} \bigr|^{2}
    = V_{10}^{2} + 2 \bigl| V_{11} \bigr|^{2}
    = x^{2} + 2 ( \alpha^{2} + \beta^{2} )
    \leq \frac{1}{2},
\label{qub3} \end{equation}
which is an analog of relation \eref{blo1}. The surface
$x^{2} + 2 ( \alpha^{2} + \beta^{2} ) = 1/2$ can be represented
parametrically
\begin{equation} \fl
    \mathrm{Re} \bigl\{ V_{11} \bigr\}
    ~=~ \frac{\sin t}{2} \cos u,
    \qquad
    \mathrm{Im} \bigl\{ V_{11} \bigr\}
    ~=~ \frac{\sin t}{2} \sin u,
    \qquad
    V_{10}
    ~=~ \frac{\cos t}{\sqrt{2}},
\label{qub4} \end{equation}
where $t \in [ 0, \pi ]$  and $u \in [ 0, 2 \pi ]$. This is
a prolate spheroid which in the present case corresponds to
the "standard" Bloch sphere. All allowed vectors $\wek{V}$
representing a 2-dimensional density matrix lie within this
spheroid, while pure states occupy its surface. Hence,
the proposed description of the density matrix yields
results fully equivalent to the "standard" one.

\section{Example: Qutrit }

\subsection{Construction}

In this section we discuss the next example. We investigate
the 3-level system, sometimes called a qutrit. For such a system
we have to $N=3$, and that entails  $j=1$ in the spirit of
section 3. The operator basis is spanned by 9 polarization
operators $\hopb{T}_{LM}(1)$ with $L=0,1,2$ and $M=-L,\ldots,L$
Applying the rule \eref{def} we construct the corresponding
matrices. They are as follows
\renewcommand{\arraystretch}{1.0}
\begin{displaymath}
  \hopb{T}_{00} = \frac{1}{\sqrt{3}}
  \renewcommand{\arraystretch}{1.0}
  \left( \begin{array}{rrr}
   1 & 0 & 0 \\  0 & 1 & 0 \\ 0 & 0 & 1 \end{array} \right),
\end{displaymath}
\begin{displaymath} \fl
\hopb{T}_{1+1}  = - \frac{1}{\sqrt{2}}
    \left( \begin{array}{rrr}
    0 & 1 & 0 \\   0 & 0 & 1 \\  0 & 0 & 0 \end{array} \right),
\quad
\hopb{T}_{10} = \frac{1}{\sqrt{2}}
    \left( \begin{array}{rrr}
    1 & 0 & 0 \\ 0 & 0 & 0 \\ 0 & 0 & -1 \end{array} \right),
\quad
\hopb{T}_{1-1} = \frac{1}{\sqrt{2}}
    \left( \begin{array}{rrr}
    0 & 0 & 0 \\ 1 & 0 & 0 \\ 0 & 1 & 0 \end{array} \right),
\end{displaymath}
\begin{displaymath} \fl
\hopb{T}_{2+2}
    = \left( \begin{array}{rrr}
      0 & 0 & 1 \\ 0 & 0 & 0 \\ 0 & 0 & 0 \end{array} \right),
\quad
\hopb{T}_{2+1}
    = \frac{1}{\sqrt{2}}
      \left( \begin{array}{rrr}
      0 & -1 & 0 \\ 0 &  0 & 1 \\ 0 &  0 & 0 \end{array} \right),
\quad
\hopb{T}_{20}
    = \frac{1}{\sqrt{6}}
      \left( \begin{array}{rrr}
      1 &  0 & 0 \\ 0 & -2 & 0 \\ 0 &  0 & 1 \end{array} \right),
\end{displaymath}
\begin{equation}
\hopb{T}_{2-1} = \frac{1}{\sqrt{2}}
    \left( \begin{array}{rrr}
    0 &  0 & 0 \\ 1 &  0 & 0 \\ 0 & -1 & 0 \end{array} \right),
\quad
\hopb{T}_{2-2} =
    \left( \begin{array}{rrr}
    0 & 0 & 0 \\ 0 & 0 & 0 \\ 1 & 0 & 0 \end{array} \right).
\label{t3d} \end{equation}
Thus, as expected, matrix $\hopb{T}_{00}$ is proportional
to the identity one, and the corresponding density operator
is written, according to the prescriptions \eref{rhot1} and
\eref{rhot2}, as
\begin{equation}
    \hat{\rho}
    = \frac{1}{3} \: \id_{3}
    + \sum_{L=1}^{2} \sum_{M=-L}^{L} V_{LM} \hopb{T}_{LM}
    = \frac{1}{3} \: \id_{3} + \wek{V} \cdot \wek{\hopb{T}},
\label{qro1} \end{equation}
where $\wek{V} = \bigl( V_{1-1}, V_{10}, V_{11}, V_{2-2},
V_{2-1}, V_{20}, V_{21}, V_{22} \bigr)$ (ordered as written)
is a generalized Bloch vector representing a qutrit. Moreover,
we note that according to relation \eref{voo} we can also
write $V_{00} = 1/\sqrt{3}$. Eight complex components $V_{LM}$
must satisfy relations \eref{vherm}. As it follows from the
comments given after these relations they are also specified
by 8 real numbers. Since dealing with real quantities is simpler
and more transparent, we introduce the following notation
\begin{eqnarray}
   V_{10} &= x,
   & \qquad V_{11} = - V_{1-1}^{\ast} = a + i b,
\nonumber \\
   V_{20} &= y,
   & \qquad V_{21} = - V_{2-1}^{\ast} = \alpha_{1} + i \beta_{1},
     \qquad V_{22} = V_{2-2}^{\ast} = \alpha_{2} + i \beta_{2},
\label{qtaa} \end{eqnarray}
with $x,a,b,y,\alpha_{1},\beta_{1},\alpha_{2},\beta_{2} \in
\mathbb{R}$. With the aid of this notation we can explicitly
write down the the density matrix for a qutrit
\renewcommand{\arraystretch}{2.5}
\begin{equation}
\arraycolsep=4mm \fl
   \hat{\rho}
   = \left( \begin{array}{ccc}
       \displaystyle{\frac{1}{3}}
     + \displaystyle{\frac{x}{\sqrt{2}}}
     + \displaystyle{\frac{y}{\sqrt{6}}}
   & - \displaystyle{\frac{a+ib}{\sqrt{2}}}
     - \displaystyle{\frac{\alpha_{1}+i\beta_{1}}{\sqrt{2}}}
   & \alpha_{2}+i\beta_{2}
   \\
     - \displaystyle{\frac{a-ib}{\sqrt{2}}}
     - \displaystyle{\frac{\alpha_{1}-i\beta_{1}}{\sqrt{2}}}
   &   \displaystyle{\frac{1}{3}}
        -\displaystyle{\frac{2}{\sqrt{6}}}\:y
   & - \displaystyle{\frac{a+ib}{\sqrt{2}}}
     + \displaystyle{\frac{\alpha_{1}+i\beta_{1}}{\sqrt{2}}}
   \\
       \alpha_{2}-i\beta_{2}
   & - \displaystyle{\frac{a-ib}{\sqrt{2}}}
     + \displaystyle{\frac{\alpha_{1}-i\beta_{1}}{\sqrt{2}}}
   &   \displaystyle{\frac{1}{3}}
     - \displaystyle{\frac{x}{\sqrt{2}}}
     + \displaystyle{\frac{y}{\sqrt{6}}}
     \end{array} \right),
\renewcommand{\arraystretch}{1.0}
\label{qtro} \end{equation}
which is clearly hermitian and normalized, as it should be.

To identify matrix \eref{qtro} as a true density matrix
we must be sure that it is positive. The positivity conditions
correspond to the inequalities $S^{(3)}_{2} \geq 0$ and
$S^{(3)}_{3} \geq 0$. The first one, as already discussed,
is equivalent to
\begin{equation}
    0 \leq \bigl| \wek{V} \bigr|^{2} \leq \frac{2}{3},
\label{posa} \end{equation}
with the left inequality being trivial due to the definition
of $|\wek{V}|^{2}$. $S^{(3)}_{3}$ follows immediately from
\eref{sknb} and it reads
\begin{equation}
    S^{(3)}_{3}
    = \frac{1}{3} \left[ \: \frac{1}{9}
    - \frac{1}{2} \: \bigl| \wek{V} \bigr|^{\:2}
    + T_{3} \right] \geq 0.
\label{posb} \end{equation}
The length $|\wek{V}^{2}|$ of the generalized Bloch vector
follows from the definition \eref{vvsp} and from
identifications \eref{qtaa}. It is given as
\begin{equation}
    \bigl| \wek{V} \bigr|^{2}
    = x^{2} + y^{2} + 2 \bigl( a^{2} + b^{2}
      + \alpha_{1}^{2} + \beta_{1}^{2}
      + \alpha_{2}^{2} + \beta_{2}^{2}\bigr),
\label{qtv2} \end{equation}
The next quantity necessary to investigate the positivity
of the density matrix \eref{qtro} is $T_{3} = \Tr \{ ( %
\wek{V} \cdot \wek{\hopb{T}} )^{3} \}$. It follows from
relations \eref{pohh} and \eref{trnna}. Computation of the
triple trace is a bit tedious but otherwise straightforward
since the CGC are tabulated in \cite{var}.
So, with identifications \eref{qtaa} we obtain
\begin{eqnarray}
 & \fl
  T_{3} = ~\frac{3x^{2}y - y^{3}}{\sqrt{6}}
     + 3 \sqrt{2} \; x \bigl( a \alpha_{1} + b \beta_{1} \bigr)
     + \sqrt{\frac{3}{2}} \; y
      \bigl( 2 \alpha_{2}^{2} + 2 \beta_{2}^{2} - a^{2} -b^{2}
             - \alpha_{1}^{2} - \beta_{1}^{2}  \bigr)
\nonumber \\[3mm]
  & + 3 \left[ \alpha_{2} \bigl( a^{2} - b^{2} - \alpha_{1}^{2}
             + \beta_{1}^{2} \bigr)
        + 2 \beta_{2} \bigl( ab - \alpha_{1}\beta_{1} \bigr)
        \right]
\label{qtt3} \end{eqnarray}
Both quantities $| \wek{V} |^{2}$ and $T_{3}$ are real, as they
should be. Using relations \eref{qtv2} and \eref{qtt3}
one expresses $S^{(3)}_{3}$ via the introduced 8 real
variables. Moreover, one easily checks that $S^{(3)}_{3} = %
\det \hat{\rho}$, as is should be.

\subsection{Parametrization with two nonzero variables}

General analytical discussion of positivity conditions
\eref{posa} and \eref{posb} together with \eref{qtv2}
and \eref{qtt3} seems to be extremely difficult if not
virtually impossible, because there are 8 real parameters.
Therefore, we will restrict our attention to a simpler case.
Namely, we will assume that only two of real parameters
$(x,y,a,b,\alpha_{1},\beta_{1},\alpha_{2},\beta_{2})$
are nonzero while the other six ones are put to zero.
Similar procedure was employed by Kimura \cite{kim} and
Byrd with Khaneja \cite{bkh}. Then, there are 28 different
pairs of nonzero parameters. We shall show that these
28 pairs split into 7 distinct types.

We shall denote the pair of nonzero parameters by $(s,t)$
and next we will indicate to which pairs taken from the set
$(x,y,a,b,\alpha_{1},\beta_{1},\alpha_{2},\beta_{2})$ it
corresponds. Then we will give then conditions \eref{posa}
and \eref{posb} written in terms of parameters $s,t$.
The formal expressions for these conditions are common to all
representatives of the given type. To discuss \eref{posb}
we introduce the quantity
\begin{equation}
    F =\frac{1}{9} - \frac{1}{2} \: \bigl| \wek{V} \bigr|^{\:2}
        + T_{3} \geq 0
\label{qtf} \end{equation}
which will be a function of two variables $s$ and $t$
and the inequality obviously follows from \eref{posb}.
After presenting the basic properties of each of the seven
types we will briefly state the properties which are common
to all of them.

\subsubsection{Type I}

\begin{figure}[ht] \begin{center}
\scalebox{0.47}[0.47]{\includegraphics{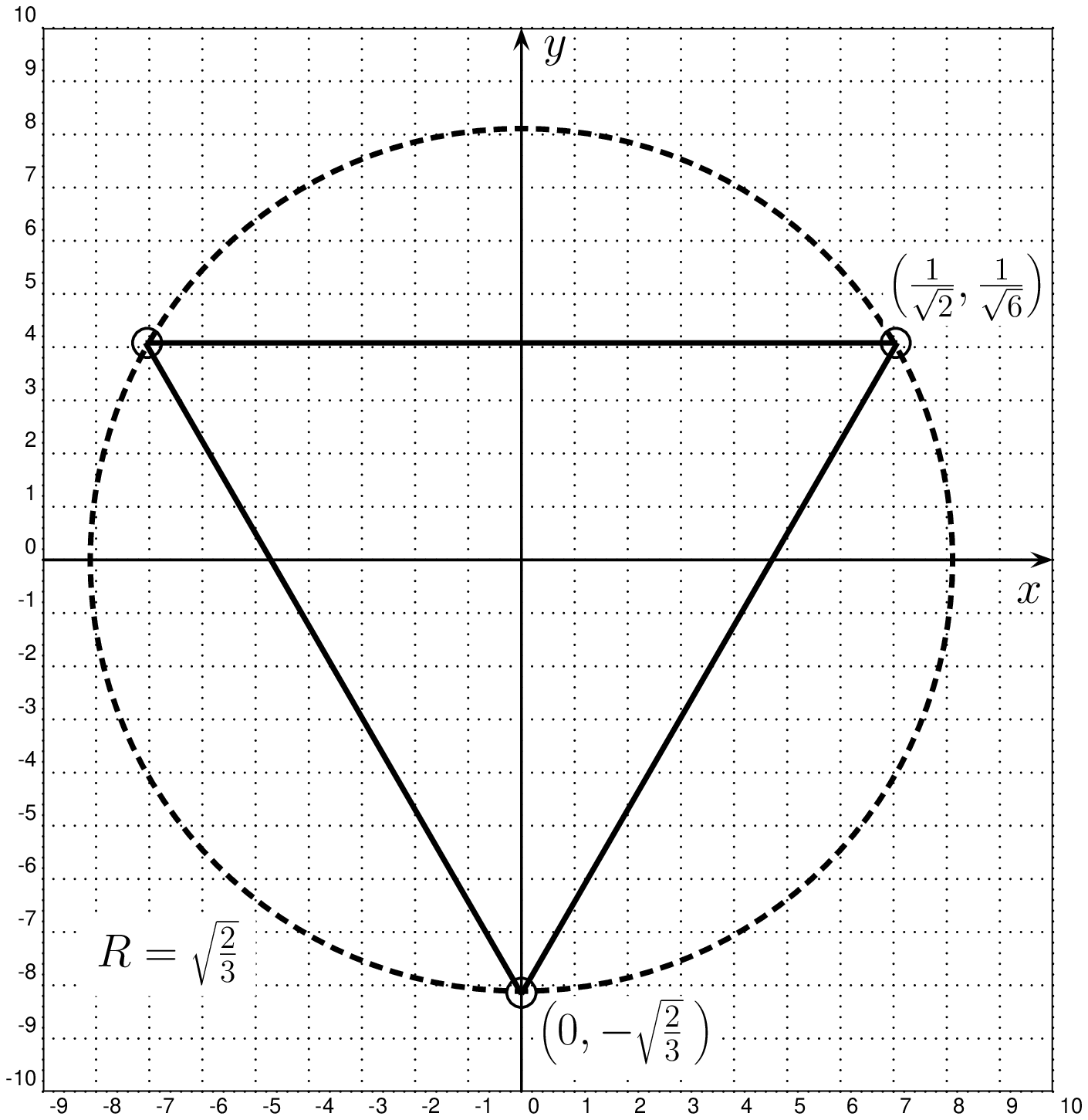}}
\scalebox{0.54}[0.54]{\includegraphics{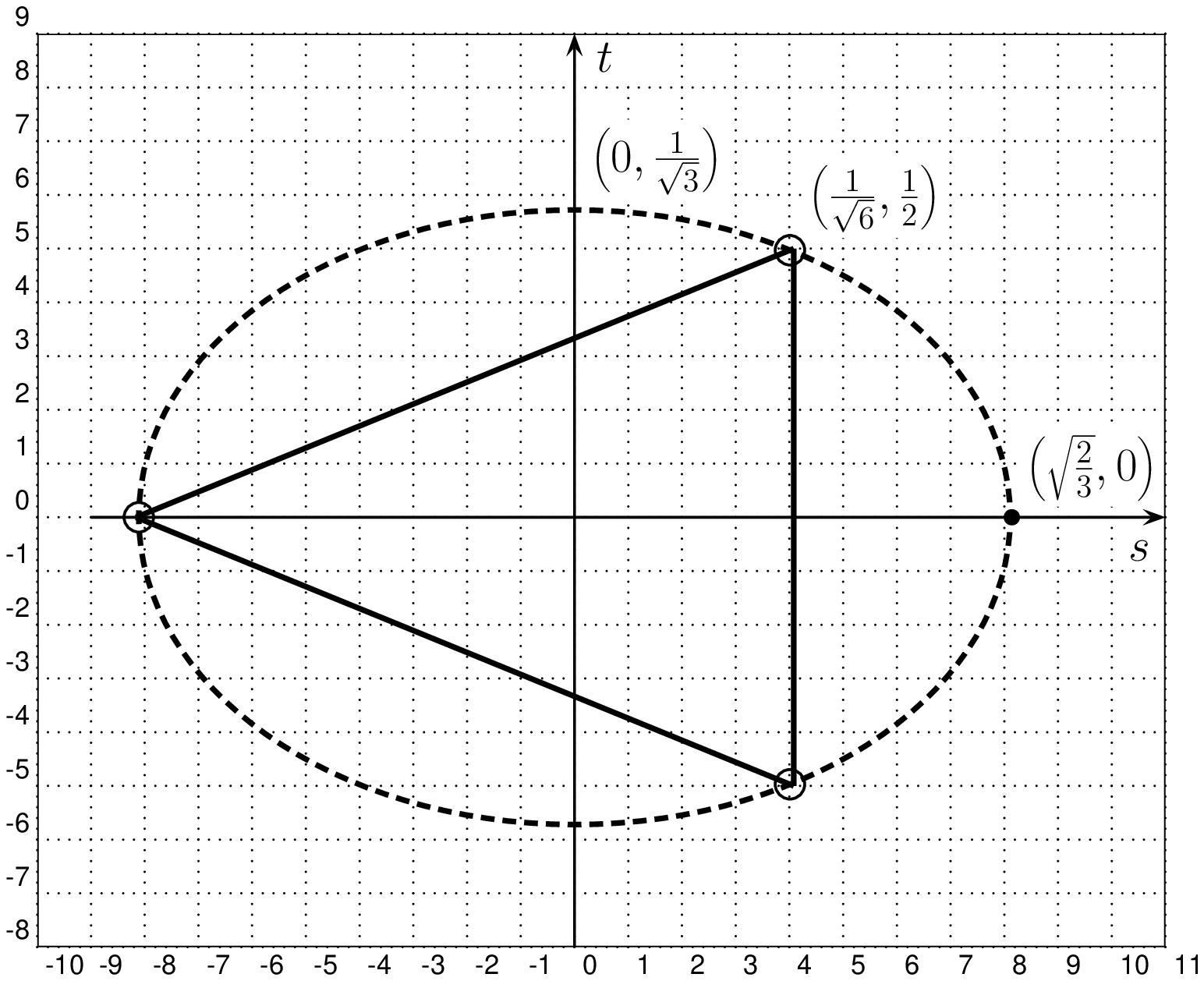}}
\caption{Type I (a) left and type II (b ) right. Allowed
states lie on the solid lines and inside the triangles.
Small circles -- pure states. The grid unit corresponds
to actual value equal to $0.1$.}
\end{center} \end{figure}

Type I corresponds to only one pair of parameters, that is
to $(s,t) = (x,y)$. Then, requirements \eref{posa} and
\eref{qtf} translate into the following ones
\numparts
\begin{eqnarray}
   \bigl| \wek{V} \bigr|^{2}
    = x^{2} + y^{2} \leq \frac{2}{3},
\label{typ1a} \\
   F(x,y) = \frac{1}{9}
   - \frac{x^{2}+y^{2}}{2}
   + \frac{3x^{2}y - y^{3}}{\sqrt{6}} \geq 0.
\label{typ1b} \end{eqnarray} \endnumparts
Requirement \eref{typ1a} restricts allowed values of $x$, $y$
to the circle of the radius $R=\sqrt{2/3}$ (dashed line).
Then \eref{typ1b} implies that the allowed points must lie
within or on the triangle indicated by solid lines in
Figure 1. We also note that for this type there are three
possible pure states $(\pm1/\sqrt{2},1/\sqrt{6})$ and
$(0,-\sqrt{2/3})$. They are indicated by small circles
in Figure 1.

\subsubsection{Type II}

Type II has two representatives $(s,t) = (y,\alpha_{2}), %
(y,\beta_{2})$. Then relations \eref{posa} and \eref{qtf}
give
\numparts
\begin{eqnarray}
   \bigl| \wek{V} \bigr|^{2}
   = s^{2} + 2 \,t^{2} \leq \frac{2}{3},
\label{typ2a} \\
   F(s,t) = \frac{1}{9} - \frac{s^{2}}{2} - t^{2}
   + \frac{6 t^{2}s - s^{3}}{\sqrt{6}} \geq 0.
\label{typ2b} \end{eqnarray} \endnumparts
Inequality \eref{typ2a}  places the allowed values of parameters
on and inside the ellipse with semi-axes of lengths
$\sqrt{2/3}$ and $\sqrt{1/3}$ (dashed line in Figure 1 (right)).
Inequality \eref{typ2b} restricts $s$ and $t$ to the triangle
drawn with solid lines in Figure 1 (right). There are also
three pure states $(\sqrt{1/6},\pm \sqrt{1/2})$
and $(-\sqrt{2/3},0)$.

\subsubsection{Type III}

\begin{figure}[ht] \begin{center}
\scalebox{0.68}[0.68]{\includegraphics{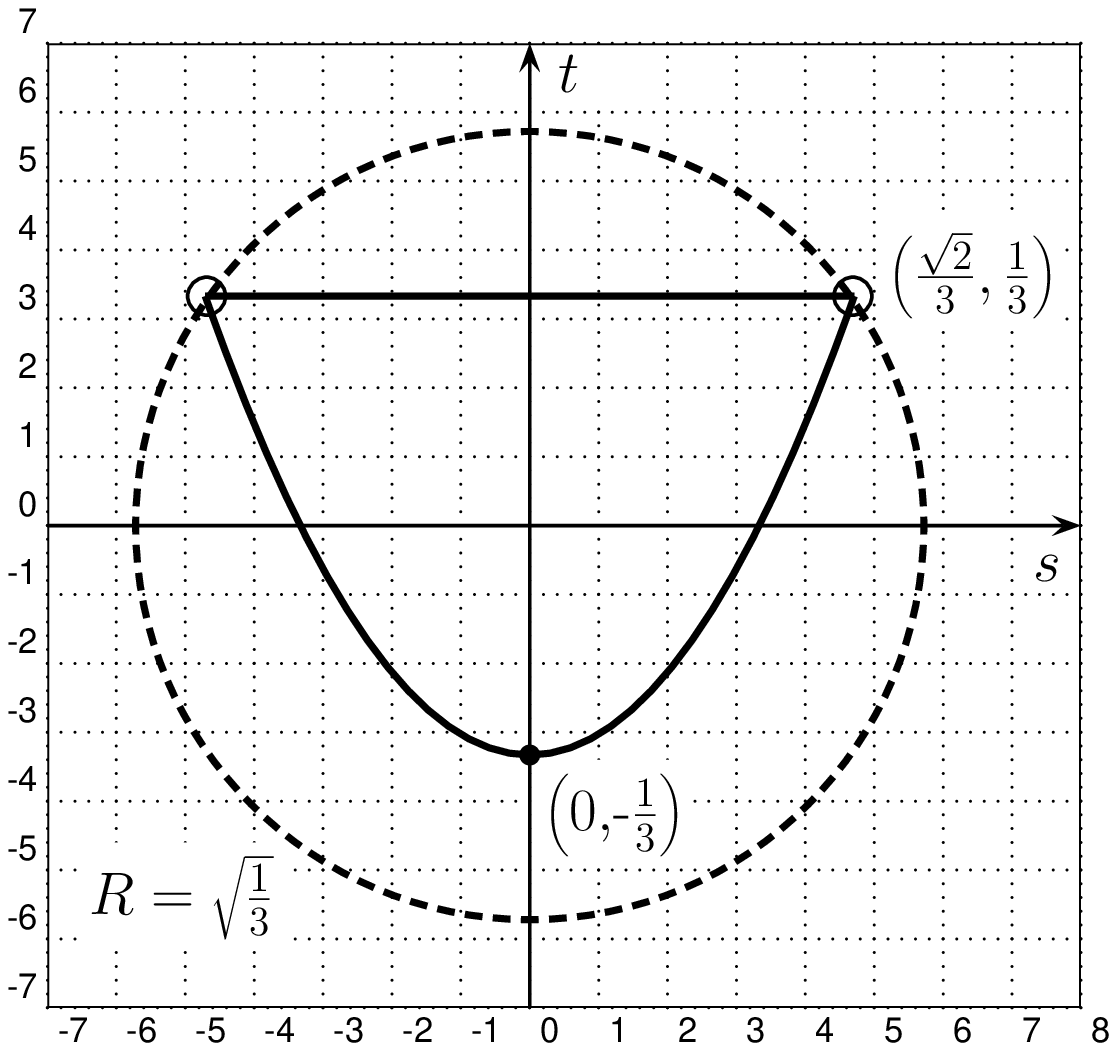}}
\scalebox{0.68}[0.68]{\includegraphics{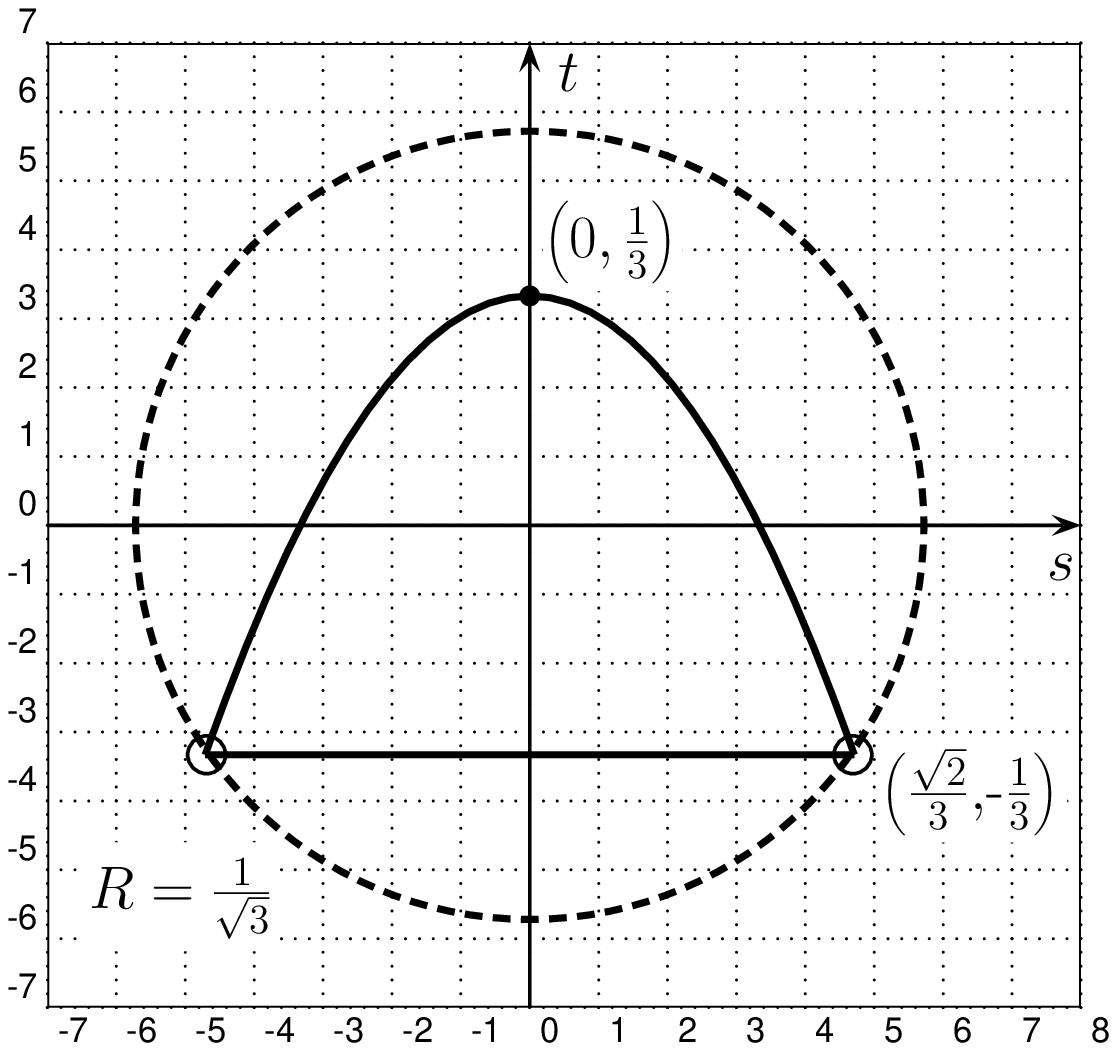}}
\caption{Type III (a) left and type IV (b) right. Allowed
states lie on the solid lines and inside the region restricted
by a horizontal straight lines and parabolas. Small circles --
pure states. The grid unit, as previously, corresponds
to actual value equal to $0.1$}
\end{center} \end{figure}

Type III is specified by two cases $(s,t) = (a,\alpha_{2}), %
(\beta_{1},\alpha_{2})$. Then from \eref{posa} and \eref{qtf}
we get
\numparts
\begin{eqnarray}
   \bigl| \wek{V} \bigr|^{2}
   = 2 ( s^{2} +  t^{2} )  \leq \frac{2}{3},
\label{typ3a} \\
   F(s,t) = \frac{1}{9} - s^{2} - t^{2}
   + 3 s^{2} t \geq 0.
\label{typ3b} \end{eqnarray} \endnumparts
The first condition gives allowed values of parameters $s$
and $t$ within a circle of radius $R=\sqrt{1/3}$ (dashed line
in Figure 2 (left). The requirement \eref{typ3b} restricts
the values of $s$, $t$ to a region below the straight line
$t=1/3$ and above the parabola $t=3s^{2} -1/3$. This region
is indicated by a solid line in Figure 2 (left). This type
allows for two pure states $(\pm \sqrt{2}/3,1/3)$ which are
denoted by small circles.

\subsubsection{Type IV}

Type IV is similar to the previous one. In this case we also have
two cases $(s,t) = (b,\alpha_{2}), (\alpha_{1},\alpha_{2})$.
Then from \eref{posa} and \eref{qtf} it follows that
\numparts
\begin{eqnarray}
   \bigl| \wek{V} \bigr|^{2}
   = 2 ( s^{2} +  t^{2} )  \leq \frac{2}{3},
\label{typ4a} \\
   F(s,t) = \frac{1}{9} - s^{2} - t^{2}
   - 3 s^{2} t \geq 0.
\label{typ4b} \end{eqnarray} \endnumparts
The imposed conditions are thus similar. Only now, inequality
\eref{typ4b} implies that the allowed points are above the
straight line $t= - 1/3$ and below the parabola
$t = - 3s^{2} +1/3$ drawn as solid lines in Figure 2 (right).
Two pure states correspond to points $(\pm \sqrt{2}/3, -1/3)$.

\subsubsection{Type V}

\begin{figure}[ht] \begin{center}
\scalebox{0.6}[0.6]{\includegraphics{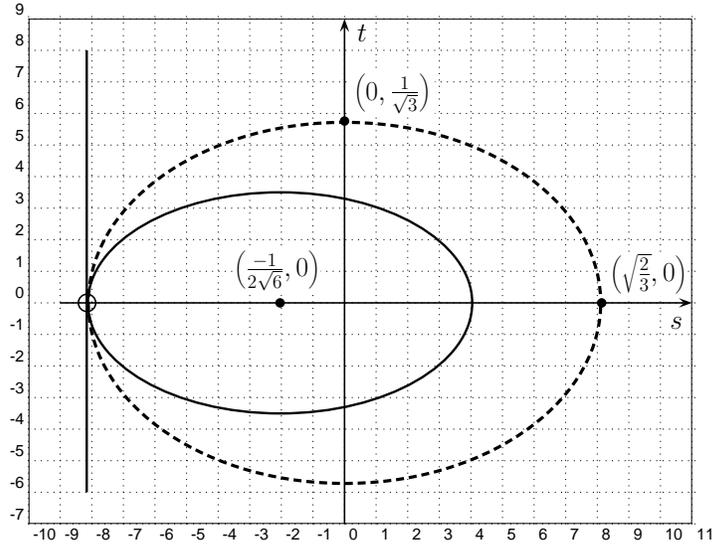}}
\caption{Type V. Allowed states lie on and inside the solid
ellipse. Small circles -- pure states. The grid unit equals
to $0.1$}
\end{center} \end{figure}

Type V has four representatives, namely
$(s,t) = (y,a), (y,b), (y,\alpha_{1}),$ $(y,\beta_{1})$.
Next, from \eref{posa} with \eref{qtf} we have
\numparts
\begin{eqnarray}
   \bigl| \wek{V} \bigr|^{2}
   = s^{2} + 2 t^{2}  \leq \frac{2}{3},
\label{typ5a} \\
   F(s,t) = \frac{1}{9} - \frac{s^{2}}{2} - t^{2}
   - \frac{3 t^{2} s + s^{3}}{\sqrt{6}} \geq 0.
\label{typ5b} \end{eqnarray} \endnumparts
The first requirement puts the allowed points within an
ellipse with semi-axes of lengths $\sqrt{2/3}$ and $\sqrt{1/3}$.
Condition \eref{typ5b} implies that the values of $s$ and $t$
are to the right of the straight line $s=-\sqrt{2/3}$ and within
an ellipse the center of which is shifted to the point
$(-1/2\sqrt{6},0)$. Semi-axes of this ellipse are equal to
$\sqrt{3/8}$ and $\sqrt{1/8}$. So, the allowed points are
within (and on) the smaller ellipse which is drawn with a
solid line in Figure 3. Here, there is only one pure state
at the point $(-\sqrt{2/3},0)$ indicated by a small circle.

\subsubsection{Type VI}

\begin{figure}[ht] \begin{center}
\scalebox{0.6}[0.6]{\includegraphics{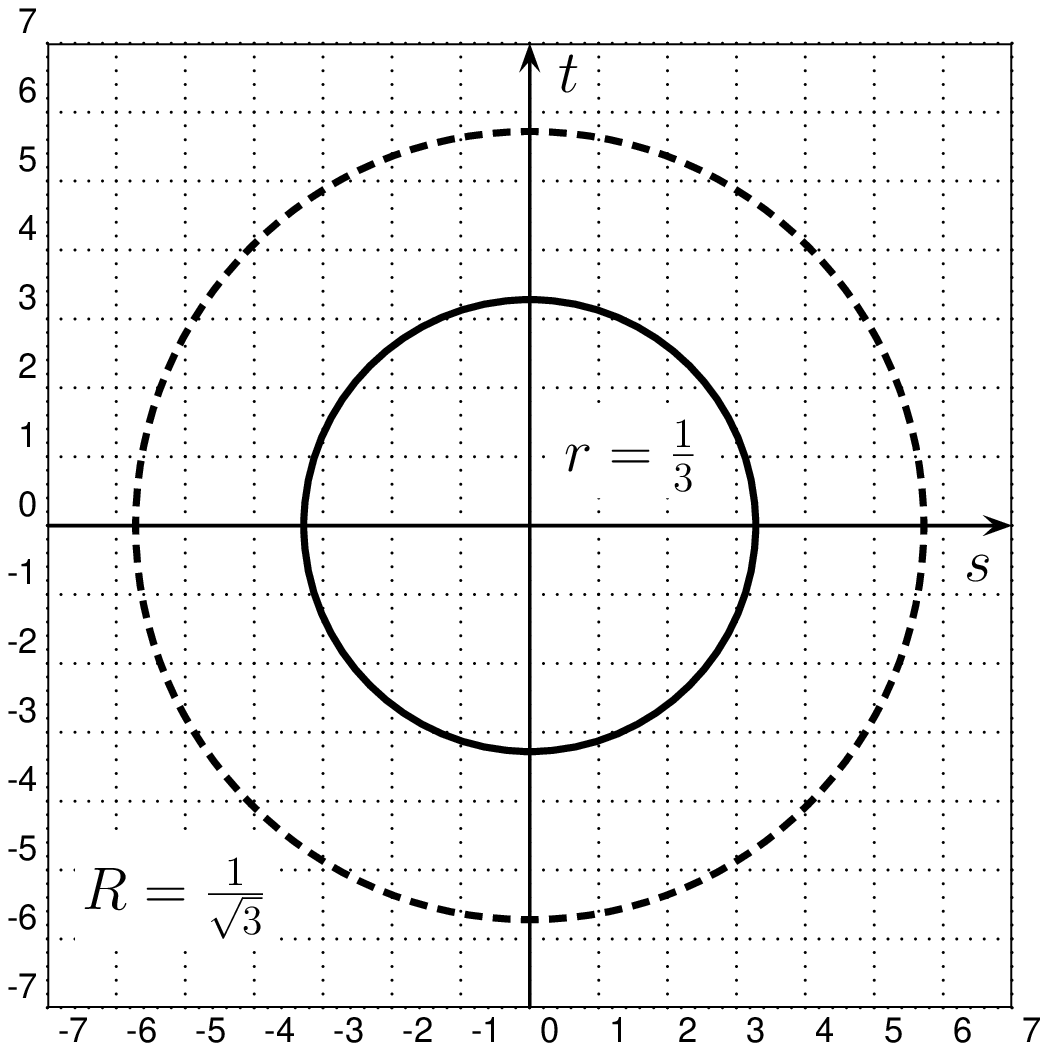}}
\scalebox{0.5}[0.5]{\includegraphics{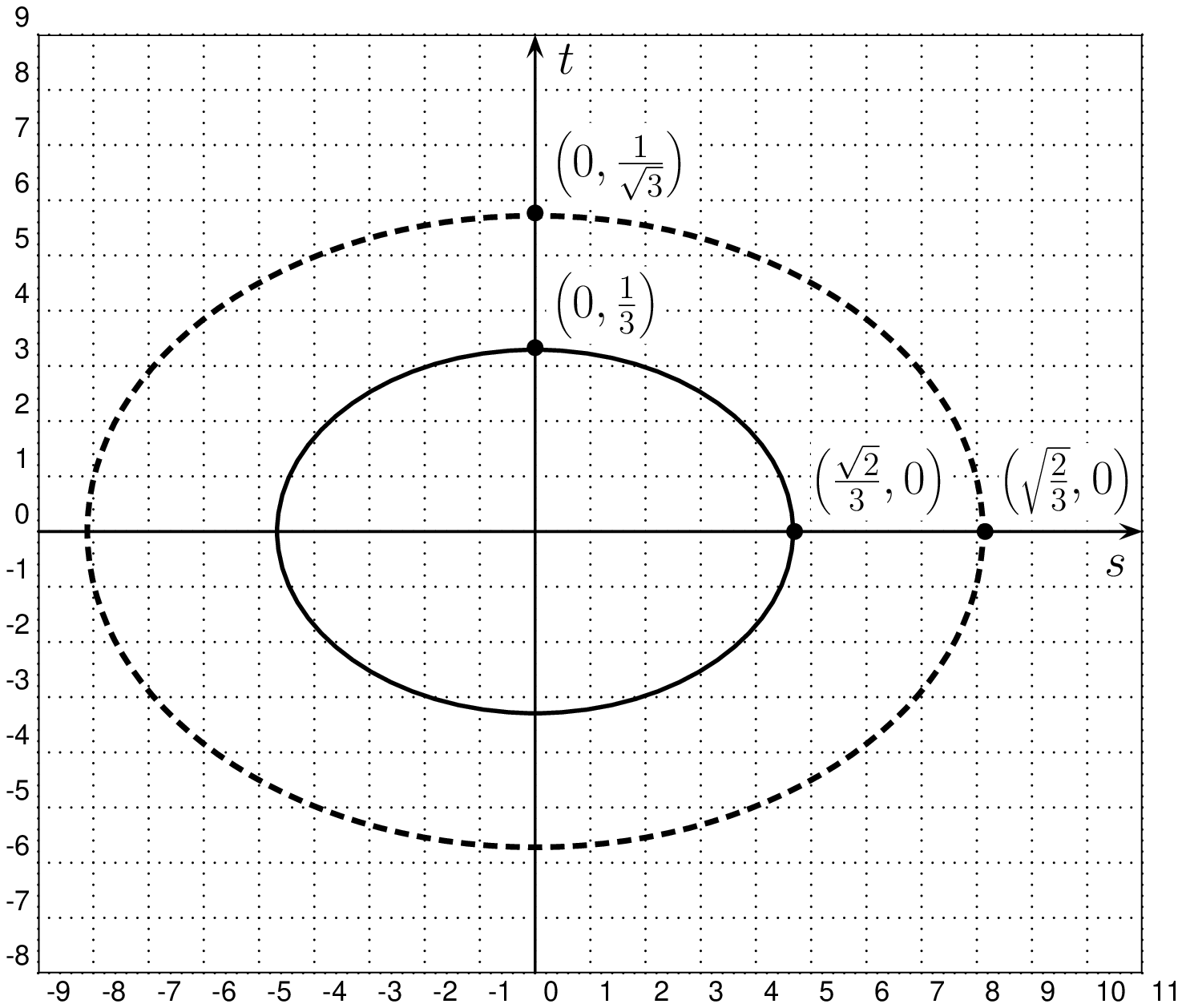}}
\caption{Type VI (left) and type VII (right). The allowed
states lie on and within the solid curves. There are no
pure states. The grid unit equals $0.1$, as always.}
\end{center} \end{figure}

This type is most numerous, it is specified by 11 cases.
They are $(s,t) = (a,b), (a,\alpha_{1}), (a,\beta_{1}),%
(a,\beta_{2}), (b,\alpha_{1}), (b,\beta_{1}), (b,\beta_{2}),%
(\alpha_{1},\beta_{1}), (\alpha_{1},\beta_{2}), %
(\beta_{1},\beta_{2}),$ $(\alpha_{2},\beta_{2})$.
Then, requirements \eref{posa} and \eref{qtf} yield
\numparts
\begin{eqnarray}
   \bigl| \wek{V} \bigr|^{2}
   = 2 ( s^{2} + t^{2} ) \leq \frac{2}{3},
\label{typ6a} \\
   F(s,t) = \frac{1}{9} - s^{2} - t^{2} \geq 0,
\label{typ6b} \end{eqnarray} \endnumparts
because in this case we find $T_{3}=0$. Both conditions specify
circles. Inequality \eref{typ6a} gives a circle with radius
$R=\sqrt{1/3}$ while \eref{typ6b} yields $r=\sqrt{1/3}$.
The smaller circle (more restrictive) is drawn solid in
Figure 4. No pure states are allowed here.

\subsubsection{type VII}

The last type is given by 6 pairs, that is by
$(s,t) = (x,a),$  $(x,b), (x,\alpha_{1}),$
$(x,\beta_{1}), (x,\alpha_{2}),$ $(x,\beta_{2})$.
Then relations \eref{posa} and \eref{qtf} give the
requirements
\numparts
\begin{eqnarray}
   \bigl| \wek{V} \bigr|^{2}
   =  s^{2} + 2 t^{2} \leq \frac{2}{3},
\label{typ7a} \\
   F(s,t) = \frac{1}{9} - \frac{s^{2}}{2} - t^{2} \geq 0,
\label{typ7b} \end{eqnarray} \endnumparts
since in this case we also have $T_{3}=0$.
Condition \eref{typ7a} restricts the allowed values of $s$
and $t$ to an ellipse with semi-axes equal to $\sqrt{2/3}$
and  $\sqrt{1/3}$. The second condition restricts the
allowed points to within and on an ellipse with
semi-axes $\sqrt{2}/3$ and $1/3$. The smaller ellipse
is drawn solid in Figure 4. As in the previous type
there are no pure states.

\subsection{Features common to all seven types}

All discussed types are characterized by two-parameter
behaviour of $|\wek{V}|^{2}$ and $F(s,t)$, the latter
being proportional to $\det \hat{\rho}$ (up to a factor
of $1/3$). In all considered cases requirement $F(s,t) \geq 0$
is more restrictive than the condition imposed upon the
length of the generalized Bloch vector $\wek{V}$.
That confirms the idea that the set of all density matrices
is a proper subset of the hyperball determined by
requiring that $|\wek{V}|^{2} \leq 2/3$.

The allowed values of parameters lie on and inside the
solid contours which correspond to $F(s,t) = 0$. Since for
pure states we have $|\wek{V}|^{2}=2/3$ and
$\det\hat{\rho}=0$ it is not surprising  that pure states
are situated at the extremal points of solid contours.

Inside these contours $F(s,t)$ is obviously positive and
for all cases attains its maximal value equal to $1/9$
at the point $(0,0)$ which corresponds to $\wek{V}=0$,
that is to a maximally mixed state.

\section{Final remarks}

We have presented and discussed the representation of the
$N \times N$ dimensional density matrix in terms of
polarization operators $\hopb{T}_{LM}(j)$. This idea is not
entirely new (see, for example \cite{var,biel}), but the
usefulness of this expansion seems to justify our recollection
of known facts. On the other hand, we have discussed the
important issue of positivity which seems not, in the
context of polarization operators, to be considered in the
literature known to us.

Usefulness of the presented approach, as it seems to us,
stems mainly from the fact that polarization operators
are expressed in terms of quantities well-known from the
quantum theory of angular momentum. Connection with this
theory greatly facilitates all considerations and allows
derivation of expressions valid for any $N=2j+1$.
For example, it is straightforward to find commutators,
products, traces over products, etc. This can be done
analytically, but also with the aid of computer programs
allowing symbolic mathematics. These possibilities seem
to indicate that the discussed approach is indeed
useful in practice.

Polarization operators $\hopb{T}_{LM}(j)$ constitute
a basis in the space of $N \times N$ dimensional
operators. Hence, the density operator can be expanded
as in \eref{rhot2}. This establishes a relationship
between $N \times N$ dimensional density matrices and
$\wek{V}$'s -- generalized Bloch vectors. Expansion
\eref{rhot2} automatically ensures proper normalization
of the density operator. This is due to the fact that
polarization operators are traceless. Next, requirement
of hermiticity implies that complex components
$V_{LM}^{\ast} =(-1)^{M}V_{L-M}$ which reduces the
number of independent real parameters to $N^{2}-1$
and fixes the dimension of the space of generalized
Bloch vectors.

Density operator must be not only normalized and hermitian
but also positive. This problem seems not to be discussed
earlier a in terms of polarization operators. We have done
that and presented the formalism allowing one to check
whether the positivity conditions are met by a given
hermitian and normalized matrix. Successive positivity
requirements (that is inequalities $S^{(N)}_{k} \geq 0$
for $k=2,3,\ldots,N$) are derived for any $N$.
These requirements specify and restrict the space of
generalized Bloch vectors which represent true density
matrices. Expressions \eref{pohh} together with
\eref{trnna} allow computation of the quantities
$T_{m} = \Tr \{ (\wek{V} \cdot \wek{\hopb{T}} )^{m}\}$
which, in turn, are used to compute coefficients
$S^{(N)}_{k}$ as in \eref{skff}. In particular,
requirement $S^{(N)}_{2} \geq 0$ entails
$\Tr\{\hat{\rho}^{2}\} \leq 1$ which, in terms of
Bloch vector yields $|\wek{V}|^{2} \leq (N-1)/N$.
So, all vectors $\wek{V}$ lie within a $N^{2}-1$
dimensional hyperball of radius equal to $\sqrt{(N-1)/N}$.
However, further conditions $S^{(N)}_{k} \geq 0$
(for $k \geq 3$) severely restrict the set of allowed
Bloch vectors. This set is a proper subset of the mentioned
hyperball and it possesses quite a complicated structure.
In the two previous sections we employed the presented
procedure to a qubit ($N=2$) and to qutrit ($N=3$).
The former is quite simple. On the other hand analysis
of a qutrit gives support to all above given remarks.

The employed representation of the density operator seems
to be quite useful. We feel that we have shown that
it can successfully be used to investigate issues which
were not studied earlier and which seem to be of contemporary
interest. Moreover, the procedure to investigate positivity
is valid for any $N$.

Finally, we would like to indicate some possibilities which
probably deserve further attention. Polarization operators
are spherical irreducible tensors \cite{var}. Therefore,
their tensor products preserve their character. This fact
seems to be promising in the investigations of entangled
states. We hope that this can provide new insights into
the structure and geometry of entangled states.
Moreover, as indicated by Biederharn and Louck \cite{biel},
polarization operators consitute just another
(non-standard) set of $SU(N)$ generators. This set is
endowed with an interesting stucture as it follows from
careful inspection of the structure constants in \eref{cctt}.
Investigation based upon this fact go beyond the scope
of this work, but seem to be an interesting subject for
future studies of the space of allowed generalized Bloch
vectors and therefore of the geometry of the space of
density operators.

We end this paper with the hope that the revitalized
expansion of the density operator in terms of polarization
operators will prove useful in other investigations.
The presented discussion of the correspondence between
density operators and generalized Bloch vectors (especally
in the light of requirement of positivity) is applicable
to any dimension. The quantities appearing here are
strongly connected with angular momentum theory and
thereby well-known. This, in our minds, indicates that
the representation discussed here may be more practical
then the "standard" one.



\section*{References}


\begin{thebibliography}{99}
\small
\bibitem{keyl}
    Keyl M 2002
    {\it Phys. Rep.} {\bf 369} 431--548
\bibitem{kim}
    Kimura G 2002
    {\it Phys. Let. A} {\bf 314} 339--349
\bibitem{bkh}
    Byrd M S Khaneja N 2003
    {\it Phys. Rev. A} {\bf 68} 062322
\bibitem{jaks}
    Jak{\'{o}}bczyk L Siennicki M 2001
    {\it Phys. Lett. A} {\bf 286} 383-390
\bibitem{mix}
    Jaegger G Sergienko A V Saleh B E A and
    Teich M C 2003
    {\it Phys. Rev. A} {\bf 68} 022318
\bibitem{gant} Gantmacher F R 1959 1960 1977 {\it Theory of Matrices}
    (New York: Chelsea Publishing Company)
\bibitem{var}
    Varshalovich D A Moskalev A N  and Khersonskii W K 1975
    {\it Quantum Theory of Angular Momentum}
    (Leningrad: Nauka, Leningrad)
    (Russian version is available tu us) \\
    Varshalovich D A Moskalev A N  and Khersonskii W K 1988
    {\it Quantum Theory of Angular Momentum}
    (Singapore: World Scientific)
\bibitem{biel}
    Biedenharn L C Louck J D 1981 {\it Angular momentum
    in Quantum Physics. Theory and Application}
    (Boston: Addison-Wesley, Reading)
\bibitem{ts1}
    Tilma T Sudarshan ECG 2002
    \JPA {\bf 35} 10467--501
\bibitem{alen}
    Alicki R Lendi K 1987 {\it Quantum Dynamical Semigroups
    and Applications} (New York: Springer)
\bibitem{cohen}
    Cohen-Tannoudji C Dupont-Roc J and Grynberg G 1992
    {\it Atom-Photon Interactions} (New York: Wiley)
\bibitem{puri}
    Puri, R R 2001 {\it Mathematical Methods of Quantum Optics}
    (New York: Springer)
\end{thebibliography}
\end{document}